\documentclass[twocolumn]{aastex631}

\usepackage{enumitem}
\usepackage{amsmath}
\usepackage{mathrsfs}
\usepackage{comment}
\usepackage{hyperref}
\usepackage{fontawesome5}

\newcommand{\github}[1]{%
   \href{#1}{\faGithubSquare}%
}

\shortauthors{Sasseville et al.}
\graphicspath{{./}{figures/}}

\begin{document}

\title{A novel approach to understanding the link between supermassive black holes and host galaxies}

\author[0000-0001-8845-2025]{Gabriel Sasseville}
\affiliation{Department of Physics, University of Montreal, Montreal, QC H2V 0B3, Canada}
\affiliation{Ciela - Montreal Institute for Astrophysical Data Analysis and Machine Learning, Montreal, QC H2V 0B3, Canada}
\affiliation{Centre de recherche en astrophysique du Québec (CRAQ), Canada}


\author[0000-0001-7271-7340]{Julie Hlavacek-Larrondo}
\affiliation{Department of Physics, University of Montreal, Montreal, QC H2V 0B3, Canada}
\affiliation{Ciela - Montreal Institute for Astrophysical Data Analysis and Machine Learning, Montreal, QC H2V 0B3, Canada}
\affiliation{Centre de recherche en astrophysique du Québec (CRAQ), Canada}

\author[0000-0001-7549-5560]{Samantha C. Berek}
\affiliation{Dunlap Institute for Astronomy \& Astrophysics, University of Toronto, 50 St George Street, Toronto, ON M5S 3H4, Canada}
\affiliation{Data Sciences Institute, University of Toronto, 17th Floor, Ontario Power Building, 700 University Ave, Toronto, ON M5G 1Z5, Canada}

\author[0000-0003-3734-8177]{Gwendolyn M. Eadie}
\affiliation{Dunlap Institute for Astronomy \& Astrophysics, University of Toronto, 50 St George Street, Toronto, ON M5S 3H4, Canada}
\affiliation{Data Sciences Institute, University of Toronto, 17th Floor, Ontario Power Building, 700 University Ave, Toronto, ON M5G 1Z5, Canada}
\affiliation{Department of Statistical Sciences, University of Toronto, 9th Floor, Ontario Power Building, 700 University Avenue, Toronto, ON M5G 1Z5, Canada}

\author[0000-0003-2001-1076]{Carter Lee Rhea}
\affiliation{Department of Physics, University of Montreal, Montreal, QC H2V 0B3, Canada}
\affiliation{Ciela - Montreal Institute for Astrophysical Data Analysis and Machine Learning, Montreal, QC H2V 0B3, Canada}
\affiliation{Centre de recherche en astrophysique du Québec (CRAQ), Canada}

\author[0000-0001-7179-9049]{Aaron Springford}
\affiliation{Toronto, Ontario, Canada}

\author[0000-0003-4440-259X]{Mar Mezcua}
\affiliation{Institute of Space Sciences (ICE, CSIC), Campus UAB, Carrer de Magrans, 08193 Barcelona, Spain}
\affiliation{Institut d'Estudis Espacials de Catalunya (IEEC), Edifici RDIT, Campus UPC, 08860 Castelldefels (Barcelona), Spain}

\author[0000-0001-6803-2138]{Daryl Haggard}
\affiliation{Ciela - Montreal Institute for Astrophysical Data Analysis and Machine Learning, Montreal, QC H2V 0B3, Canada}
\affiliation{Centre de recherche en astrophysique du Québec (CRAQ), Canada}
\affiliation{Department of Physics, McGill University, 3600 rue University, Montréal, QC H3A2T8, Canada}
\affiliation{Trottier Space Institute at McGill, 3550 rue University, Montréal, QC H3A2A7, Canada}



\begin{abstract}

The strongest and most universal scaling relation between a supermassive black hole and its host galaxy is known as the $M_\bullet-\sigma$ relation, where $M_\bullet$ is the mass of the central black hole and $\sigma$ is the stellar velocity dispersion of the host galaxy. This relation has been studied for decades and is crucial for estimating black hole masses of distant galaxies. However, recent studies suggest the potential absence of central black holes in some galaxies, and a significant portion of current data only provides upper limits for the mass. Here, we introduce a novel approach using a \textit{Bayesian hurdle model} to analyze the $M_\bullet-\sigma$ relation across 244 galaxies. This model integrates upper mass limits and the likelihood of hosting a central black hole, combining logistic regression for black hole hosting probability with a linear regression of mass on $\sigma$. From the logistic regression, we find that galaxies with a velocity dispersion of $11$, $34$ and $126$ km/s have a $50$\%, $90$\% and $99$\% probability of hosting a central black hole, respectively. Furthermore, from the linear regression portion of the model, we find that $M_\bullet \propto \sigma^{5.8}$, which is significantly steeper than the slope reported in earlier studies. Our model also predicts a population of under-massive black holes ($M_\bullet=10-10^5 M_\odot$) in galaxies with $\sigma \lesssim 127$ km/s and over-massive black holes ($M_\bullet \geq 1.8 \times 10^7$) above this threshold. This reveals an unexpected abundance of galaxies with intermediate-mass and ultramassive black holes, accessible to next-generation telescopes like the Extremely Large Telescope.

\end{abstract}



\section{Introduction} \label{sec:intro}

The stellar velocity dispersion ($\sigma$) of the bulge of a galaxy correlates strongly with the mass ($M_\bullet$) of the supermassive black hole (BH) at its center. This relation, known as the $M_\bullet-\sigma$ relation, has been observed empirically for decades (e.g. \citealt{merritt99}; \citealt{ferrarese00}; \citealt{gultekin09a}; \citealt{thater19}; \citealt{dullo21}) and takes the form of $M_\bullet \propto \sigma^{5.4}$ (e.g. \citealt{bosch16}) for galaxies in the local Universe. It is considered a dynamical scaling relation. The mere existence of this correlation suggests coevolution between supermassive BHs and their host galaxies (e.g. \citealt{kormendy13a}). Indeed, the region within which the gravitational potential of the BH dominates the gravitational potential of the host galaxy, known as the sphere of influence, is significantly smaller (e.g. $\approx 0.8$ parsec for a BH with mass $10^6 M_\odot$) than the size of its host galaxy which extends to dozens of kpc (e.g. \citealt{merritt09}, \citealt{kormendy13a}). The use of this dynamical scaling relation has proven to be crucial in estimating BH masses in distant galaxies at $z > 0.1$ (e.g. \citealt{salviander13}), for which it is difficult or impossible to obtain high resolution data for direct BH mass measurements with current telescopes.

Although the scatter in the $M_\bullet-\sigma$ relation is low ($0.40$ dex) for high-mass BHs with $M_\bullet\geq10^6 M_\odot$, it does not succeed in fitting the data at the low-mass end, where it rather follows $M_\bullet \propto \sigma^{3.32}$ (e.g. \citealt{xiao11}). Thus, little is known about the transition region that lies between high and low velocity dispersion galaxies. This region corresponds to the range of BH masses for which there is a change in the slope of the $M_\bullet-\sigma$ relation (e.g. \citealt{martin-navarro18}).
Furthermore, some observational data suggest the lack of a central BH in some galaxies (e.g. \citealt{merritt01a}, \citealt{gebhardt01}, \citealt{valluri05}, \citealt{coccato06}, \citealt{lyubenova13}, \citealt{nguyen14}), a phenomenon which is not predicted by the $M_\bullet-\sigma$ relation. The lack of BHs in these galaxies, however, could be due to observational limits. For example, some dwarf galaxies thought to lack a central BH were recently found to have active galactic nuclei (AGN) once observational and estimation techniques improved (e.g. \citealt{mezcua18b}, \citealt{baldassare20}, \citealt{mezcua20}, \citealt{reines22}). Thus, we cannot say with certainty that some galaxies lack a central BH, and instead have upper limits on BH mass.

A better understanding of the transition mass as well as a detailed comparison of the low-mass and high-mass regions of the $M_\bullet-\sigma$ relation could shed light on the BH-galaxy coevolution and the possible lack of a central BH in some dwarf galaxies. More precisely, the extension of the $M_\bullet-\sigma$ relation to the low-mass regime could imply that the growth regulation in dwarf galaxies is driven by BH feedback, rather than supernova feedback (e.g. \citealt{efstathiou00}, \citealt{stinson09}, \citealt{dashyan20}, \citealt{barai18}, \citealt{baldassare20}).

Various methods have been used to model the $M_\bullet-\sigma$ relation in previous studies, such as the bivariate linear regression routine from \cite{akritas96} used in \cite{ferrarese00} or the linear regression routines from \cite{kelly07} used in \cite{bosch16}. Whilst \cite{ferrarese00} included uncertainties on both predictor and response variables (independent and dependent variables, respectively), they performed the regression on a sample of only 12 supermassive BHs where upper limits were excluded. On the other hand, \cite{bosch16} included upper limits and uncertainties on BH mass measurements, but their model was strictly linear in the log space and did not account for the potential lack of BHs in dwarf galaxies. The potential lack of a central BH was considered in \cite{gultekin09a}. They treated the probability of having no central BH as a constant value $P_\emptyset$ and included it directly in the likelihood of their maximum likelihood estimation (MLE) technique for fitting (see appendix A of \citealt{gultekin09a}). The problem with this assumption is that it may not hold for higher velocity dispersion galaxies, as it may be easier for them to form BHs (e.g. \citealt{haidar22}).

More recently, a novel fitting technique was used in \cite{eadie22} and \cite{berek23} to model the relationship between a globular cluster (GC) system's mass and its host galaxy's stellar mass, that allowed for the inclusion of zero GC counts in some galaxies. This model, known as a \textit{Bayesian hurdle model}, which we will refer to as the hurdle model (e.g. \citealt{burkner17}), combines both binary and continuous data for the best fit. The hurdle model simultaneously modelled both (1) the probability of a galaxy containing a GC system, and (2) the linear relation between the mass of that system and the host galaxy's stellar mass. They found that this technique was able to accurately model the data and provide them with new insights on the transition mass region ($10^5 - 10^8 M_\odot$) where galaxies go from having no GC system to having one. Their findings suggest that the mass of GCs within this region are subject to large scatter and that the least massive GCs are likely to disappear due to dynamical evolution.

Here, we apply a similar hurdle model to the $M_\bullet - \sigma$ relation. Our reasons for doing this are twofold: (1) the data from a statistical perspective is strikingly similar to that in \cite{eadie22} and \cite{berek23}, and (2) the hurdle model overcomes many of the aforementioned issues of fitting the $M_\bullet - \sigma$ relation. The novelty of the method is that it allows us to take into account systems that may not have a BH, without assuming a constant probability for the BH occupation. We also modify the hurdle model to include measurement uncertainties on BH mass measurements as well as uncertainties on upper limits. This task is not trivial as the uncertainties on the upper limits are heteroskedastic, but could provide valuable insight regarding this scaling relation (e.g. \citealt{eadie22}).

Our results show that considering upper limits and uncertainties on BH mass measurements tends to drag the high-mass end of the curve upwards and the low-mass end of the curve downwards while maintaining a similar trend for the transition region. If correct, the new behavior of the $M_\bullet - \sigma$ relation at the low-mass and high-mass end would have significant implications for our understanding of the coevolution between BHs and galaxies. 

This paper is structured as follows. In Section \ref{sec:data}, we present the observational data. In Section \ref{sec:model and methods}, we describe the hurdle model and its application to our dataset. Afterwards, we consolidate both binary and continuous distributions by applying this model to our sample. In Section \ref{sec:discussion}, possible physical interpretations of our results as well as a comparison with previous results are presented. We finish with Section \ref{sec:conclusion}, in which we summarize our findings.

Throughout this paper, a flat concordance cosmology with $H_0 = 70$ km/s and $\Omega_m = 0.3$ is adopted.

\section{Observational sample} \label{sec:data}

Our sample is comprised of 244 galaxies for which the central BH masses have been dynamically measured or temporally resolved (see \citealt{baldassare15}, \citealt{baldassare16},  \citealt{bosch16}, \citealt{krajnovic18}, \citealt{thater19} and \citealt{baldassare20}). All BH masses along with their uncertainties are listed in Tables \ref{table:included} and \ref{table:excluded}. The details of how we compiled these data will be explained next.

The BH mass measurements are done using four different methods: gas dynamics, stellar dynamics, megamasers and reverberation mapping. The first method relies on measuring gas dynamics at the galactic center, more precisely the rotational velocity of ionized gas surrounding the BH. This measurement can then be separated into its two sources, the gravitational potential from the central BH and the extended matter distribution of the galaxy's stellar component (see \citealt{kormendy13a}, \citealt{yoon17}). The second method is based on resolving stellar kinematics near the galactic center, rather than gas dynamics (see \citealt{siopis09}, \citealt{mcconnell12}, \citealt{kormendy13a}).
The third method uses the rotation curves of $H_2O$ megamaser disks formed in some AGN. These rotation curves follow a Keplerian law, which makes it relatively simple to determine the central BH mass (see \citealt{gao16}, \citealt{greene16}, \citealt{kuo18}). The final method relies on estimating the radius of the broad-line region by measuring time lags between the AGN's continuum fluctuations. With this radius, the central BH mass can be found via the Virial theorem, assuming that the motion of the gas in the broad-line region is dominated by the BH's influence (see \citealt{bentz10}; \citealt{grier13a}; \citealt{baldassare20}). 

Our data sample includes a large number of upper limits in addition to robust measurements, but does not include lower limits on BH mass measurements. The diversity of our sample highlights the importance of including measurement uncertainties in the model. The various methods have different systematic uncertainties; megamasers have the smallest uncertainty (e.g. \citealt{gao16}, \citealt{kuo18}) and reverberation mapping is the most uncertain, as it relies on assuming the broad-line region's geometry (e.g. \citealt{grier13a}).

Our compilation is comprised of a total of $244$ BHs with 2MASS growth curves and reliable velocity dispersions at redshifts up to $z \approx 0.1$. An additional $64$ BHs are compiled in Table \ref{table:excluded}, but are omitted from the fits due to various reasons such as the lack of 2MASS growth curves, the lack of a measured velocity dispersion or due to them being significant outliers in the Fundamental Plane (FP) of BH accretion (see \citealt{merloni03}, \citealt{vdb16}, \citealt{gultekin19}). The 2MASS growth curves are essential because they provide global photometric properties necessary for evaluating BH masses and galaxy properties in the context of the Virial Theorem and the FP (e.g. \citealt{bosch16}). Furthermore, the removal of significant outliers (galaxies NGC2139 and NGC4636) to the FP helps eliminate potential biases that may stem from systematic measurement errors, ensuring a more reliable analysis (e.g. \citealt{bosch16}). Details on the omissions are discussed further in \cite{bosch16}. Including the upper limits is important because it allows us to gain insight on the low-mass end of the $M_\bullet - \sigma$ relation, considering that $14$ of the $24$ BHs with masses $\leq 10^6 M_\odot$ included in our fit are upper limits (e.g. \citealt{beifiori09}).


\subsection{BH masses} \label{subsec:BH masses}

Among the 244 BHs included in this paper, the mass measurements of 230 of them are taken from the compilation presented in \cite{bosch16}, which, in turn, were obtained from \cite{merritt01b}, \cite{gebhardt01}, \cite{valluri05}, \cite{beifiori09}, \cite{barth09}, \cite{gultekin09a}, \cite{seth10}, \cite{kormendy10}, \cite{beifiori12}, \cite{neumayer12}, \cite{mcconnell13}, \cite{kormendy13a}, \cite{lyubenova13}, \cite{scharwachter13}, \cite{de-lorenzi13}, \cite{onken14}, \cite{gultekin14}, \cite{nguyen14}, \cite{seth14}, \cite{walsh15}, \cite{yildirim15}, \cite{onishi15}, \cite{bentz15}, \cite{walsh16}, \cite{barth16b}, \cite{thomas16}, \cite{greene16}, \cite{saglia16}, \cite{onishi16}. Added to this are three BH masses from \cite{krajnovic18} and one from \cite{thater19}, all of which were measured using stellar dynamics. An additional eight BH masses from \cite{baldassare20}, one from \cite{baldassare15} and one from \cite{baldassare16} that were temporally resolved using reverberation mapping are included in the sample. This results in 244 BHs included in the regression.

We also note that BH masses are scaled by a factor $\langle f \rangle = 4.31 \pm 1.05$ if they have been measured using Virial methods (e.g. \citealt{grier13a}). Indeed, this virial factor accounts for the unknown geometry of the broad-line region, thus giving rise to a higher uncertainty than dynamical methods. Nevertheless, these methods are essential as they allow us to populate the low-mass end of our dataset. Statistical uncertainties on BH mass measurements are used when available. Otherwise, systematic uncertainties on the estimation technique are employed (e.g. $\sim 0.4$ dex for the 8 BH masses in \citealt{baldassare20}) which are usually greater than the measurement uncertainty, acting as an overestimated boundary.

We use a $3\sigma_x$ confidence level on BH mass measurements in our model, meaning the data is within 3 standard deviations ($\sigma_x$) from the mean, or rather 99.7\% of the data lies within the uncertainties, and we assume that they are Gaussian. Therefore, the uncertainties from the literature, as listed in Tables \ref{table:included} and \ref{table:excluded}, have been scaled to $3\sigma_x$ confidence levels (i.e. 99.7\%) for all regression purposes. This assumption relies on the central limit theorem (e.g. \citealt{feller1}, \citealt{billingsley_probability_1995}).

\subsection{Velocity dispersions} \label{subsec:velocity dispersions}

For all objects, we prioritize the effective stellar velocity dispersion ($\sigma_e$) over the central velocity dispersion ($\sigma_c$). This is done because $\sigma_e$ is the closest to the second moment of the 3D velocity tensor inside the half-light radius, which is the optimal measurement (e.g. \citealt{bosch16}). We use the definition for $\sigma_e$ from \cite{gultekin09a}:

\begin{equation}
    \sigma_e = \frac{ \int_{0}^{R_e} (\sigma^2 + V^2)I(r) dr }{ \int_{0}^{R_e} I(r) dr}
    \label{equation:sigma}
\end{equation}
where $I(r)$ is the surface brightness of the galaxy, $R_e$ is the effective radius and $V$ is the rotational component of the spheroid. This is done to avoid inconsistencies in the data, although \cite{gultekin09a} argues that there is no systematic bias to high or low values when comparing $\sigma_e$ to $\sigma_c$. Thus, when the effective velocity dispersion is unavailable, the central velocity dispersion from HyperLeda\footnote{See http://leda.univ-lyon1.fr/} is used. All velocity dispersions along with their uncertainties are listed in Tables \ref{table:included} and \ref{table:excluded}.

\section{Model and methods} \label{sec:model and methods}

 Hurdle models are a class of generalized linear models (GLMs; see e.g. \citealt{de-souza15b}, \citealt{elliott15}, \citealt{de-souza15a}) that allow the accurate modeling of data when zeros are present, i.e. when the response variable takes on values of zero in the dataset. Thus, they operate under the GLM framework, first introduced in \cite{nelder72}, in which a link function $g$ connects the expected value of the response $\textbf{Y}$ to the linear predictor $\textbf{X} \boldsymbol{\beta}$:
 
\begin{equation}
g(E(\textbf{Y})) = \textbf{X} \boldsymbol{\beta}.
\label{equation:GLM}
\end{equation}

 
 
Hurdle models also allow the modeling of binary and continuous data in a two-step approach, as they use two distinct parts to describe a random variable. The first portion models the probability that a  random variable, central BH mass in our case, reaches the value of zero or not using logistic regression (Section \ref{subsec:logistic regression}). The second part models the value of the random variable, using linear regression (Section \ref{sec: linear regression section}), if a nonzero value was reached in the first portion (e.g. \citealt{zuur09}). The binary portion is considered to be the 'hurdle' that must be overcome to attain a nonzero value (e.g. \citealt{hilbe17}, \citealt{eadie22}). 
 

We implement the hurdle model with the Bayesian Regression Models using Stan (\texttt{brms}) package in the R Statistical Software Environment (e.g. \citealt{burkner17}, \citealt{R}). The \texttt{brms} package automatically translates the model into the Stan language and compiles code into C++ for efficient sampling. Stan (e.g. \citealt{stan17}) uses a No-U-Turn sampler, a method introduced in \cite{hoffman11} that has been optimized to allow for efficient high-dimensional sampling (e.g. \citealt{betancourt18}). The Bayesian inference process is performed on samples approximating the posterior distribution, specifically a Markov chain of model parameters. 

The next subsections define the two part model as well as our method of including uncertainties and upper limits in the model.
 
\subsection{Logistic Regression} \label{subsec:logistic regression}
 
 
 
 
Logistic regression, a type of GLM,  operates under the framework described in Section \ref{sec:model and methods}. In logistic regression, the response for variable $i$ is assumed to follow a Bernoulli distribution with probability of success $p_i$, and probability of failure $1-p_i$ (e.g. \citealt{peacock00}). When assuming that our response variables \textbf{Y} are Bernoulli distributed, we obtain $E(\textbf{Y}) = \textbf{p}$. Combining this result with Equation (\ref{equation:GLM}), we obtain: 
 
 \begin{equation}
     \textbf{p} = g^{-1}  (\textbf{X} \boldsymbol\beta)  \\
     \label{equation:Bernoulli}
 \end{equation}
 
We use a \textit{logit} link function to map the linear predictor term $\textbf{X} \boldsymbol\beta$ from the unconstrained space $]-\infty, \infty[$ to the constrained space $[0, 1]$ (e.g. \citealt{de-souza15b}). When doing so, 
Equation (\ref{equation:Bernoulli}) becomes:

 \begin{equation}
 \textbf{p} = logit^{-1}(\textbf{X} \boldsymbol\beta) = \frac{1}{1+e^{-\textbf{X} \boldsymbol\beta}}
 \label{equation:logit}
 \end{equation}
 
 As we are only interested in the case with one predictor variable, Equation (\ref{equation:logit}) is simplified to:
 
 \begin{equation}
     p_i = \frac{1}{1+e^{-(\beta_0 + \beta_1 X_i )}} 
     \label{equation:pdef}
 \end{equation}
 
 The two parameters $\beta_0$ and $\beta_1$ are obtained by the means of 
 Bayesian inference (e.g. \citealt{albert09}, \citealt{schoot21}) where the likelihood is defined as in \cite{eadie22}:
 
 \begin{equation}
     \mathscr{L}(\boldsymbol\beta; \textbf{x}, \textbf{y}) = \prod_{i}^{N} p_i^{y_i} (1-p_i)^{1-y_i}
 \end{equation}
 
 with each $p_i$ defined by Equation (\ref{equation:pdef}).
 
 \subsection{Linear Regression}
 \label{sec: linear regression section}

Linear regression is used to model the relationship between the predictor variable and the nonzero values of the response variable resulting from the binary portion of the model. It is performed through Bayesian inference, based on the same framework as described in \cite{eadie22} and \cite{berek23}, and results in the following form:

\begin{equation}
    E[\log M_\bullet] =  \left( \gamma_0 + \gamma_1 \log\sigma \right)
    \label{equation:linearreg}
\end{equation}

where $\gamma_0$ and $\gamma_1$ are the parameters of the linear regression. 

Sections \ref{subsubsec:lognormal regression} and \ref{subsubsec: mixture_model} describe the linear regression models used for treating precise mass measurements and upper limits, respectively. The details of including uncertainties and upper limits in this portion of the model are discussed in Section \ref{sec: uncertainties and upper limits}.

\subsubsection{Lognormal Distribution}
\label{subsubsec:lognormal regression}

Our method uses the same framework as described in the previous subsection, where we use $\boldsymbol\gamma$ instead of $\boldsymbol\beta$ to avoid confusion:

\begin{equation}
 E(\textbf{Y}) = g^{-1} (\textbf{X} \boldsymbol\gamma)  \\
 \label{equation:GLM2}
 \end{equation}
 
 The canonical link function can be used to map the linear predictor term $\textbf{X} \boldsymbol\gamma$ from the unconstrained space $]-\infty, \infty[$ to the zero-truncated space $]0, \infty[$ (e.g. \citealt{hardin07}). Thus, Equation (\ref{equation:GLM2}) becomes:
 
 \begin{equation}
 E(\textbf{Y}) =  \textbf{X} \boldsymbol\gamma
 \end{equation}
 
 The assumption for the normal model described in \cite{nelder72} still holds for the lognormal model as long as the natural log of the response variable is taken prior to modeling. This assumption is as follows:
 
 \begin{equation}
     E(\textbf{Y}) = \boldsymbol\mu
 \end{equation}
 
 Combining the two previous results, we obtain the following:
 
\begin{equation}
    \boldsymbol\mu =  \textbf{X} \boldsymbol\gamma \\
    \label{eq:11}
\end{equation}

 Although the generalization to multiple predictors is quite simple, we are solely interested in the case with one predictor since we are only studying the relation between $M_\bullet$ and $\sigma$, without taking into account other factors. Thus, Equation (\ref{eq:11}) is simplified:
 
 \begin{equation}
     \boldsymbol\mu_i =  (\gamma_0 + \gamma_1 X_i) \\
 \end{equation}
 
 Once again, the two parameters $\gamma_0$ and $\gamma_1$ are obtained via Bayesian inference (e.g. \citealt{albert09}, \citealt{schoot21}) where the likelihood is defined as in \cite{hardin07}:

\begin{equation}
     \mathscr{L}(\boldsymbol\gamma; \textbf{x}, \textbf{y}) = \prod_{i}^{N} \frac{1}{\sigma\sqrt{2\pi}} \frac{1}{y_i} 
  \exp\left( -\frac{1}{2}\left(\frac{\ln(y_i)-\mu_i}{\sigma}\right)^{\!2}\,\right)
  \label{equation:lognormal}
 \end{equation}

 \subsubsection{Uncertainties and Upper Limits}
 \label{sec: uncertainties and upper limits}

The values that overcome the hurdle are typically addressed by assuming that they are lognormally distributed, as demonstrated in previous works (e.g. \citealt{eadie22}, \citealt{berek23}). In our approach, we present a novel methodology by distinguishing between precise mass measurements and nonzero upper limits, using distinct distributions for each. Specifically, we model precise mass measurements using a lognormal distribution, consistent with the approaches described in \cite{eadie22} and \cite{berek23}. 

In our treatment of upper limits, we explored several options to accurately model the uncertainty and nature of these limits, including the use of exponential and Lévy distributions. However, we found that a mixture model provided a more representative description. This model treats the upper limits as either being close to the limit or close to the established relation, allowing for more flexibility. The model uses a truncated normal distribution to represent data points near the detection limit and a lognormal distribution for values closer to the established mass relation. The weight between these two distributions, represented by the parameter $\lambda$, is learned from the data itself. This enables the model to adapt and more realistically represent cases where upper limits correspond to true masses near the relation.

\subsubsection{Mixture Model for Upper Limits}
\label{subsubsec: mixture_model}

For upper limits, we adopt a mixture model approach. The same framework and methodology from Section \ref{sec: linear regression section} applies, except the likelihood function now blends lognormal and truncated normal distributions:

\
\begin{multline}
    \mathscr{L}(\boldsymbol{\gamma},  \boldsymbol{\lambda}; \textbf{x}, \textbf{y}, \textbf{a}, \textbf{b}) = \prod_{i=1}^{N} \lambda \cdot \text{lognormal}(y_i \mid \mu_i, \sigma_i \\ + (1 - \lambda) \cdot \text{truncated\_normal}(y_i \mid \mu_i, \sigma_i, a_i, b_i)
    \label{eq: mixture_model}
\end{multline}

where $\lambda$ controls the weighting between the two distributions. The lognormal distribution is defined as in Section \ref{subsubsec:lognormal regression}, whereas the truncated normal distribution is defined as such:

\begin{equation}
     \mathscr{L}(\boldsymbol \gamma; \textbf{x}, \textbf{y}) = \prod_{i=1}^{N} \frac{1}{\sigma\sqrt{2\pi}} 
  \frac{ \exp\left( -\frac{1}{2}\left(\frac{y_i-\mu_i}{\sigma}\right)^2 \right) }{\Phi\left(\frac{b-\mu_i}{\sigma}\right) - \Phi\left(\frac{a-\mu_i}{\sigma}\right)}
  \label{equation:truncated_normal}
\end{equation}

The denominator accounts for proper normalization between the truncation limits because $\Phi(\cdot)$ is simply the cumulative distribution function of the standard normal distribution, defined as follows:

\begin{equation}
    \Phi(x) = \frac{1}{2} \left[ 1 + \text{erf} \left( \frac{x}{\sqrt{2}} \right) \right]
\end{equation}

This model allows for better handling of uncertainty in upper limits by incorporating the possibility of them representing values near the detection limit or near the established relation, thus providing a more accurate reflection of the data.




 

 \subsection{Bayesian Hurdle Model} \label{subsec:hurdle model}
 
 As mentioned before, the hurdle model combines both logistic and linear regressions. Thus, the expected response value ($log(M_\bullet)$) in terms of the predictor ($log(\sigma)$) is given by:

\begin{equation}
    E[\log M_\bullet] = \left( \frac{1}{1+e^{-(\beta_0 + \beta_1 \log\sigma)}} \right) \left( \gamma_0 + \gamma_1 \log\sigma \right)
    \label{equation:hurdle}
\end{equation}

\begin{figure*}[ht!]
\plotone{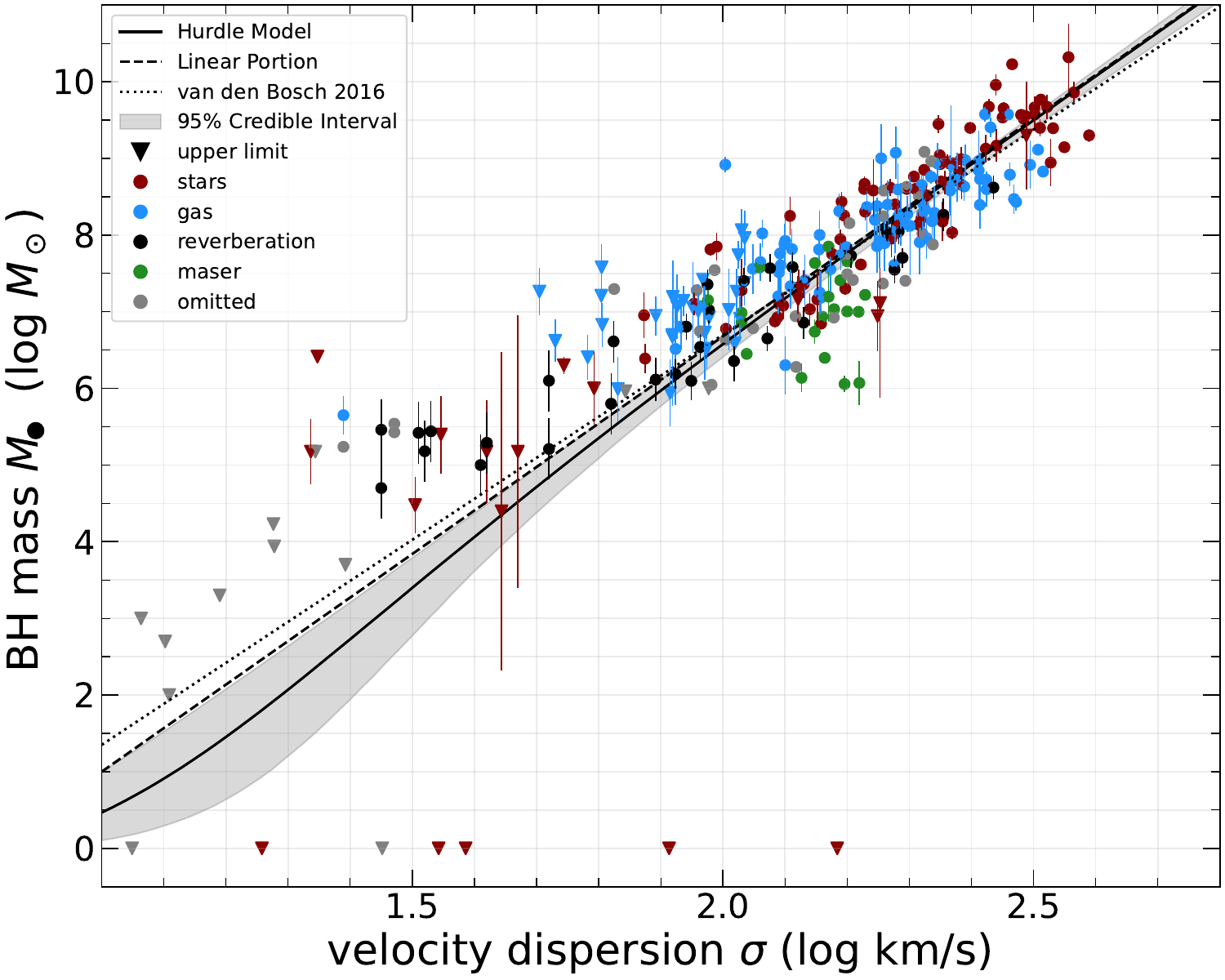}
\caption{Correlation between black hole (BH) mass and stellar velocity dispersion ($\sigma$). The solid black line shows the average fit obtained via our hurdle model, which includes uncertainties on the BH mass. The grey shaded region denotes the $95$\% Bayesian credible interval. The dashed line shows the linear portion of the fit, without considering the probability of containing a BH. The dotted black line shows the fit obtained in \cite{bosch16} with an intrinsic scatter of $\epsilon=0.49\pm0.03$. Upper limits are plotted as triangles and precise mass measurements are plotted as circles, both with their respective error bars. BH masses obtained by stellar dynamics, gas dynamics, reverberation mapping and water masers are plotted in red, blue, black and green, respectively. The BHs omitted from the fits are plotted in grey. We notice that the slope of our model's linear portion is steeper than that of \cite{bosch16}.} 
\label{fig:hurdle}
\end{figure*}

The first term is the Bernoulli probability \textbf{p} (see Equation (\ref{equation:Bernoulli})) of having a BH and it is dictated by the parameters $\beta_0$ and $\beta_1$ from the logistic regression. The second term is the linear portion of the curve and is governed by the parameters $\gamma_0$ and $\gamma_1$ from the linear regression (e.g. \citealt{eadie22}, \citealt{berek23}).

We used the \texttt{brm} function from the \texttt{brms} package to generate the base Stan code. The default prior distributions for the parameters are uniform (improper) distributions. To explore the impact of prior choices, we also considered weak (proper) priors on the fit parameters defined as follows:

\begin{align*}
    \beta_0 &\sim \mathcal{N}(0, 5) \\
    \beta_1 &\sim \mathcal{N}(5, 5) \\
    \gamma_0 &\sim \mathcal{N}(0, 5) \\
    \gamma_1 &\sim \mathcal{N}(5, 5) 
\end{align*}

Here, $\beta_0$ and $\gamma_0$ are sampled from a normal distribution with a mean $\mu=0$ and a standard deviation $\sigma=5$, while $\beta_1$ and $\gamma_1$ are sampled from a normal distribution with $\mu = 5$ and $\sigma = 5$. Prior predictive checks demonstrated that these weak priors had a minimal impact on the model results. Despite their limited influence, we chose to use these weak priors for our analysis, as they provide slightly more informative constraints compared to uniform priors.

\begin{deluxetable}{cccc}[ht]
\tablecaption{Hurdle Model Coefficients\label{tab:coefficients}}
\tablewidth{0pt}
\tablehead{
\colhead{Parameter} & \colhead{Estimate} & \colhead{95\% Credible} \\ 
\colhead{} & \colhead{}  & \colhead{Interval}}
\startdata
$\beta_0$ & -4.36  & (-8.86, 0.17)  \\
$\beta_1$ & 4.27 & (1.93, 6.87) \\
$\gamma_0$ & -4.85 & (-5.30, -4.40)  \\
$\gamma_1$ & 5.76 & (5.56, 5.95) \\
$\lambda$ & 0.98 & (0.92, 1.00) \\
\label{table:params}
\enddata
\end{deluxetable}

 Uncertainties are treated differently depending on the type of mass measurement, which is separated into three categories: (1) zero upper limits, (2) nonzero upper limits, and (3) precise measurements. The first is simply an upper limit quoted at $M_\bullet=0$, meaning there is no evidence for the presence of a BH, whereas the second is an upper limit quoted at some $M_\bullet > 0$, meaning that there is evidence for a central BH with mass equal to or lower than this upper limit, but no certainty. An example of these are the low velocity dispersion galaxies from \cite{beifiori09} and \cite{beifiori12} that present evidence of a BH, but are systematically offset from the $M_\bullet - \sigma$ relation. Thus, they are treated as nonzero upper limits. The third is a measurement for which $M_\bullet > 0$ and we are certain that there is a BH, such as NGC 1194 (e.g. \citealt{greene10a}, \citealt{greene11}).

The zero upper limits comprise $\approx 2\%$ (5/244) of the sample and are treated as zeros $(Y_i=0)$ in the logistic portion of our model. The nonzero upper limits represent $\approx 18 \%$ (44/244) of the dataset. In our two-step approach, they act as zeros ($Y_i = 0$) for the logistic portion of the model and are modeled by the mixture model in the linear portion. Their uncertainties are passed as the scale parameter in Equation (\ref{eq: mixture_model}). The remaining $\approx 80 \%$ (i.e. 195/244) of the sample consists of precise measurements and are also treated in a two-step process. They act as ones $(Y_i = 1)$ for the logistic portion and have their mass measurement modeled using the lognormal distribution in the linear portion. Their uncertainties are treated as the scale parameter $\sigma$ in Equation (\ref{equation:lognormal}).

When applying this model to our dataset, we obtain the mean parameter values shown in Table \ref{table:params}, with their respective posterior distributions in \ref{fig:param post dist}. The mean estimate is plotted as the solid black curve in Figure \ref{fig:hurdle} by using these values and Equation (\ref{equation:hurdle}). The 95\% Bayesian credible interval is shown as the grey density plot. 

We also investigated the impact of including upper limits. To achieve this, we compared the fit obtained in Figure~\ref{fig:hurdle} with a fit that disregards these upper limits. This alternative fit treats all data points as precise mass measurements, with the framework described in Section~\ref{subsubsec:lognormal regression}. The comparison is illustrated in Figure~\ref{fig:compareupplims}.

Appendix \ref{appendix:posterior} presents the posterior distributions resulting from our model. In Figure \ref{fig:param post dist}, posterior distributions of the hurdle model parameters can be seen, whereas Figure \ref{fig:bh mass post dist} presents the posterior distributions of BH mass conditioned on several values of velocity dispersion.

\section{Discussion and Comparisons} \label{sec:discussion}

\subsection{Previous studies}

As mentioned in Section \ref{sec:intro}, there have been many studies pertaining to the $M_\bullet-\sigma$ relation. One of the first was \cite{ferrarese00}, in which they used the Bivariate Corrected Errors and Intrinsic Scatter (BCES) estimator from \cite{akritas96}. The BCES estimator is intuitive because it is a generalization of the well-known ordinary least squares method. Thus, it takes the form of a simple regression model $X_{2i} = \alpha + \beta X_{1i} + e_i$, where the parameters $\alpha$ and $\beta$ are correlated to the moments of the bivariate distribution of $(X_{2}, X_{1})$ and where $e_i$ is an error term assumed to have zero mean and finite variance (e.g. \citealt{akritas96}). The BCES method is useful because it accounts for measurement errors on both predictor and response variables while also allowing for these errors to be dependent (e.g. \citealt{akritas96}). Using this method, \cite{ferrarese00} obtain $M_\bullet \propto \sigma_c^{4.8 \pm 0.5}$, which had a flatter slope than those found by recent studies. This can be explained by the more limited data at the time, resulting in a small sample size of only 12 robust BH mass measurements. Although it should not have a considerable effect on the results, we note that they used $\sigma_c$, whereas we prefer to use $\sigma_e$ when available.

\begin{figure}[ht]
\plotone{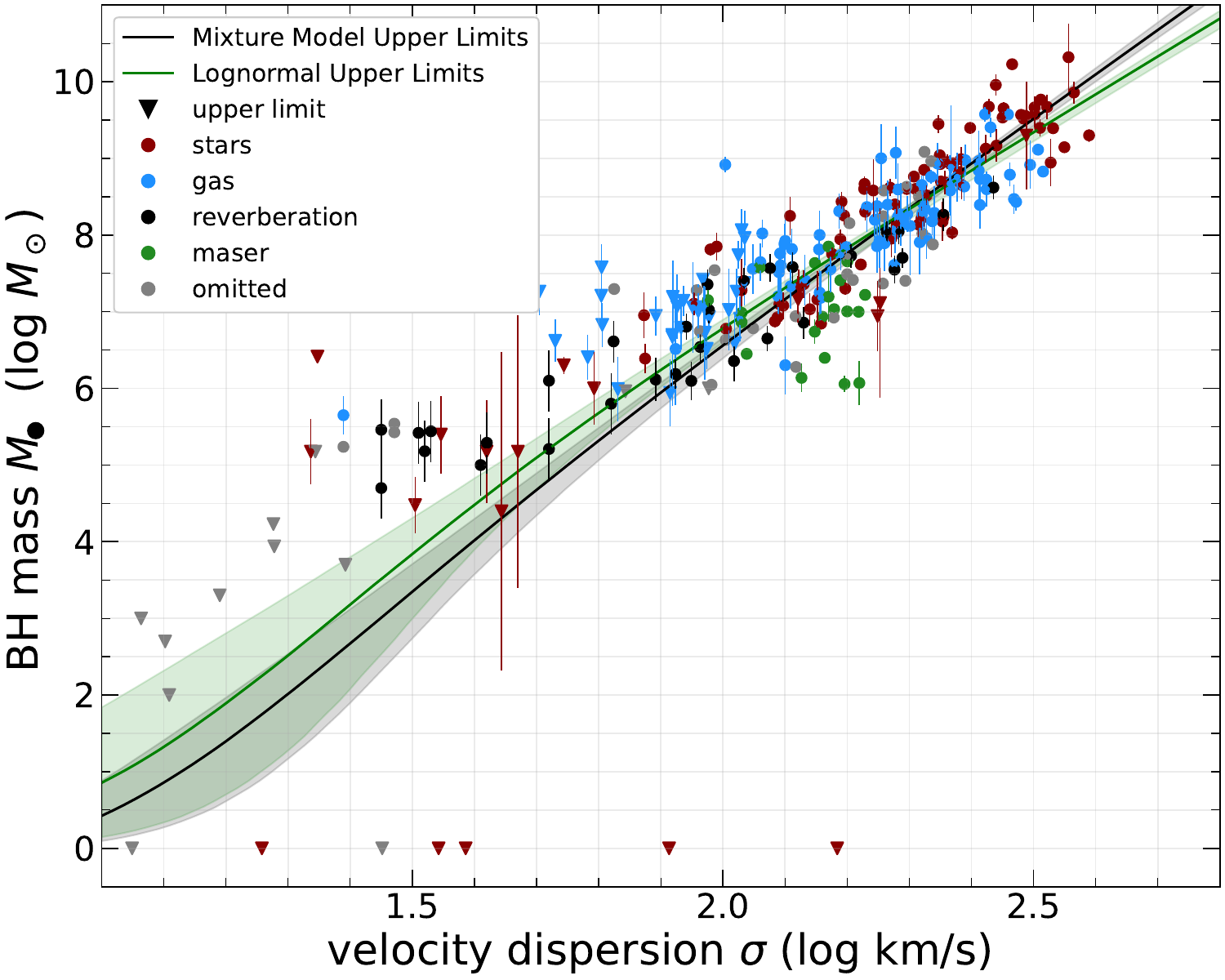}
\caption{The solid black line shows the average fit obtained, along with its $95$\% credible interval, when treating the nonzero upper limits differently than the precise mass measurements, with a mixture model rather than a lognormal likelihood. The solid green line shows the average fit obtained, along with its $95$\% credible interval, when treating the nonzero upper limits in the same manner as the precise measurements, that is with a lognormal likelihood. We notice that treating the upper limits has a considerable effect, ultimately dragging the curve downwards at the low-mass end and upwards at the high-mass end.}
\label{fig:compareupplims}
\end{figure}

Subsequently, a new fitting method introduced in \cite{gultekin09a} was used to model the $M_\bullet-\sigma$ relation. Here, a generalized MLE method that is much more complex than the BCES estimator was used. It not only includes measurement errors on both predictor and response variables, but also allows for the inclusion of upper limits on mass measurements in the model. Upper limits were included via their likelihood function which included the probability, assumed to be constant for all $\sigma$, that a galaxy may lack a central BH. Using a sample of 49 precise BH mass measurements and 19 upper limits, these authors found $M_\bullet \propto \sigma_e^{4.2 \pm 0.4}$.

The study by \cite{bosch16} encompasses the most diverse measurements as it uses a large sample of 230 BH masses, resolved via several different methods, including 49 upper limits. This will be our reference for comparison as our sample is built upon theirs. They use the \textit{mlinmixerr} and \textit{linmixerr} routines from \cite{kelly07}, which operate under a Bayesian framework to perform multiple linear regression, taking the general form $\eta_i = \alpha + \beta^T \xi_i + \epsilon_i$ (see Section 4.3 of \citealt{kelly07}). These algorithms are powerful as they account for selection effects (e.g. \citealt{leistedt16}) and nondetections in the data sample. This is particularly useful in our case because it introduces a new way of treating the upper limits. In this method, upper limits are treated as nondetections due to observational limits rather than galaxies potentially lacking a central BH. They obtain $M_\bullet \propto \sigma^{5.4 \pm 0.2}$, which is steeper than previous studies, but can be attributed to the inclusion of a larger and more diverse sample, encompassing low-mass galaxies and upper limits, in their analysis.

Treating the upper limits is a difficult task because it heavily relies on assumptions made on the nature of these nondetections. Nevertheless, it is important to include them in the $M_\bullet-\sigma$ relation as they represent a non-negligible portion of currently available data on BH masses (e.g. \citealt{boeker99}, \citealt{barth09}, \citealt{neumayer12}, \citealt{nguyen17}). Some studies treat these as nondetections due to lack of precision in the measurement methods, whereas others treat them as the lack of a central BH. 

In this paper, we account for both of these possibilities as there is significant observational evidence supporting the two theories, such as the case of M33 (e.g. \citealt{ferrarese00}, \citealt{merritt01a}, \citealt{gebhardt01}). This bulgeless galaxy has a velocity dispersion of $21$ km/s $\leq \sigma_c \leq 34$ km/s (e.g. \citealt{merritt01a}), which suggests a central BH mass of $\approx 7000$ $M_\odot$ using the relation from \cite{ferrarese00}. In \cite{gebhardt01}, they suggest a conservative upper limit on the mass of 1500 $M_\odot$, while \cite{merritt01a} suggests an upper limit of 3000 $M_\odot$. Even if the latter is true, it still lies significantly below what is predicted. However, \cite{gebhardt01} show that the Hubble Space Telescope Wide Field and Planetary Camera 2 photometry and Space Telescope Imaging Spectrograph spectroscopy data are best fit by 3-integral dynamical models when the BH mass is zero, suggesting the lack of a central BH. Other studies also suggest the lack of a BH in low-mass galaxies (e.g. \citealt{greene07}, \citealt{trump15}, \citealt{greene20}). The ambiguity surrounding whether M33 hosts a galaxy without a central BH or is simply a nondetection highlights the significance of considering both of these scenarios.

\begin{figure}[ht]
\plotone{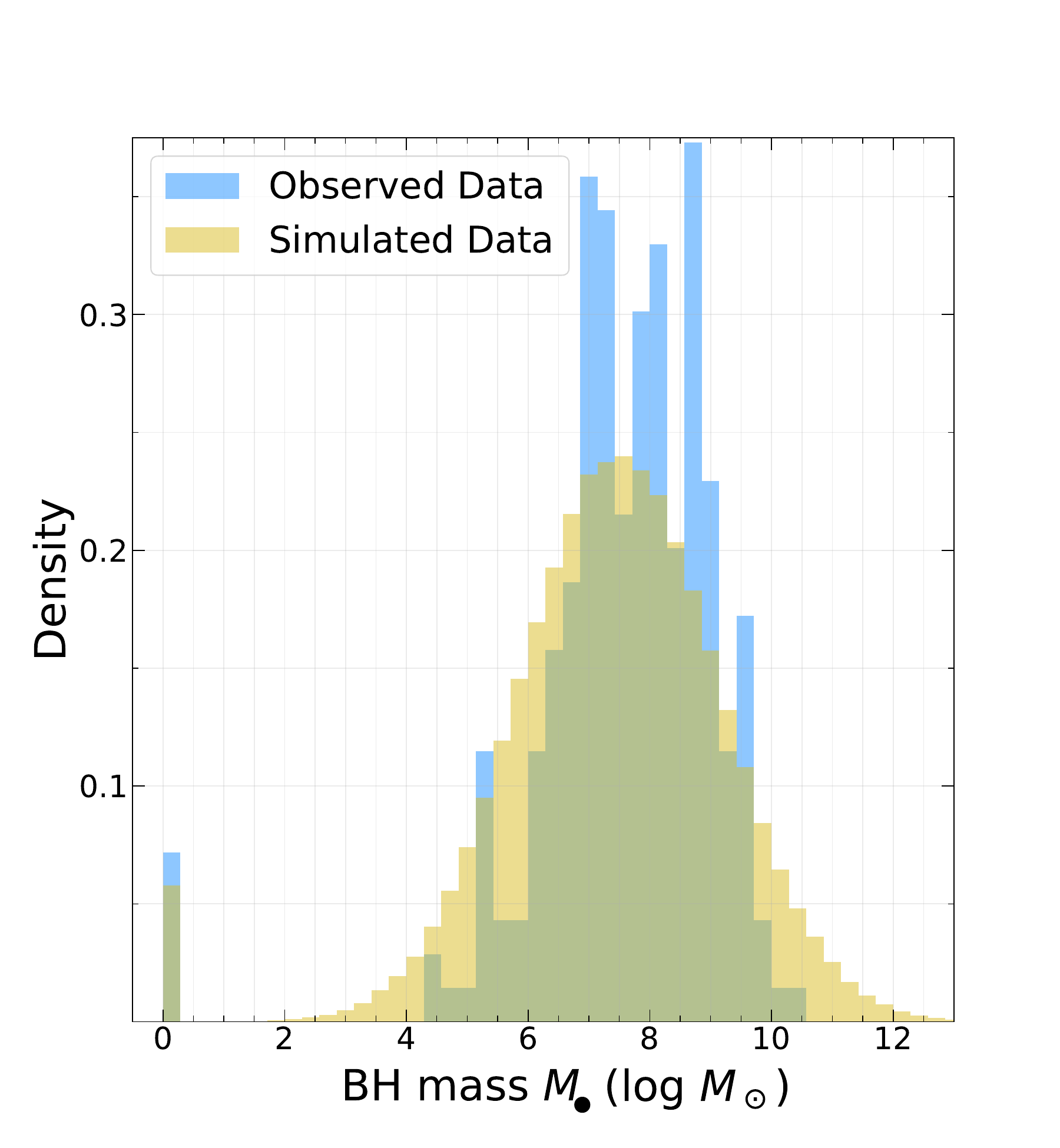}
\epsscale{0.1}
\caption{Comparison of BH mass marginal distributions between observed (blue) and simulated (yellow) data. The simulated data represents 1000 simulated points for each of the 1000 randomly chosen posterior samples. Both distributions have been normalized for comparison. Although the simulated data has much higher variance than the observed data, it manages to accurately capture the distribution of upper limits.}
\label{fig:histogram}
\end{figure}

\begin{figure*}[ht]
\plotone{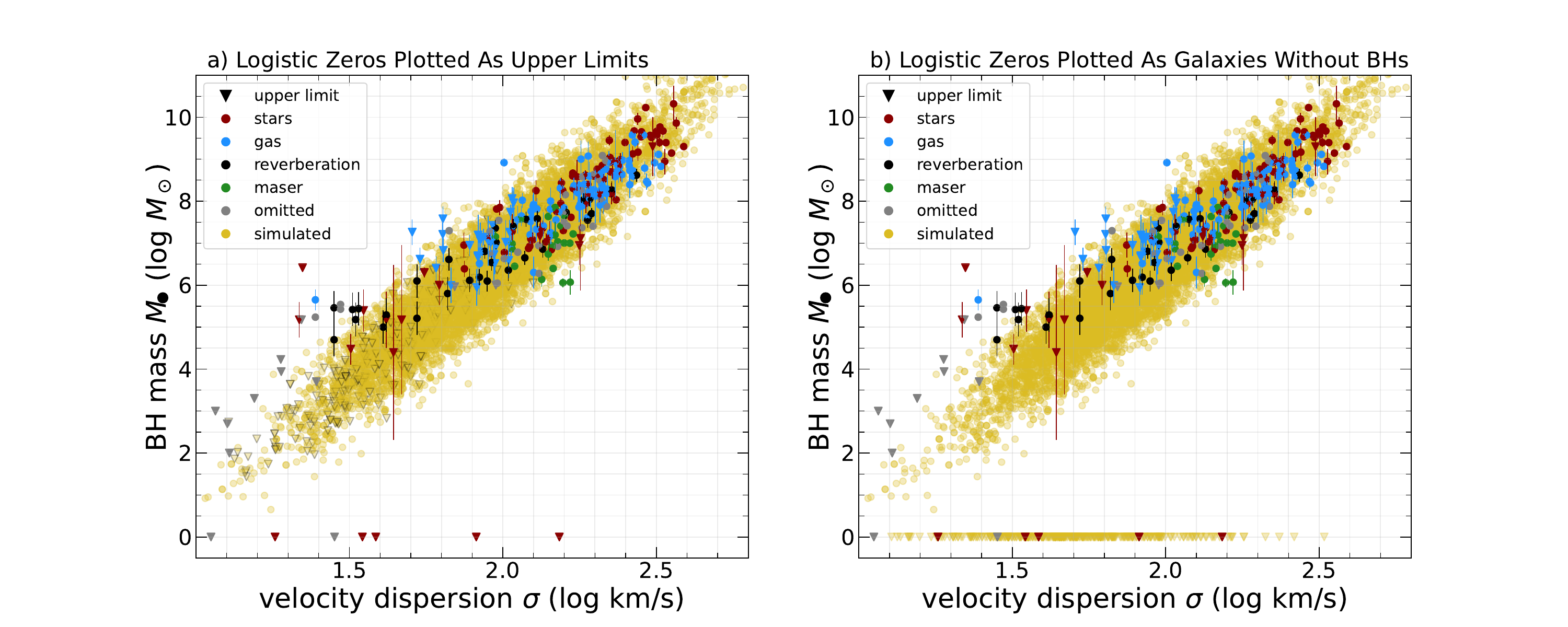}
\epsscale{0.2}
\caption{Simulated data (yellow) plotted against observational data, with two different interpretations of the logistic portion zeros: a) Treating them as upper limits, meaning galaxies that may lack a BH or are simply nondetections. b) Treating them as galaxies with no central BH. We favor the interpretation of panel a), which is more physically motivated, as discussed in Section \ref{subsubsec:postpredchecks}.}
\label{fig:postpred}
\end{figure*}

\subsection{Our results}
\label{sec:results}


We introduced a novel technique, previously used in \cite{eadie22} and \cite{berek23}, to fit the $M_\bullet-\sigma$ relation. In \cite{eadie22} and \cite{berek23}, this model was used to fit the relation between the mass of a globular cluster system and its host galaxy mass.  Their method allowed them to model the probability that a galaxy will contain a GC system, just as it allows us to estimate the probability that galaxies at a certain mass (and hence velocity dispersion) will host a BH.

The adoption of weakly informative priors, as opposed to uniform priors, had a negligible effect on our results. The median values of the linear parameters remained unchanged, while the median values of the logistic parameters, $\beta_0$ and $\beta_1$ (see Table \ref{tab:coefficients}), varied by less than $1$\%. Similarly to \cite{eadie22}, we opted for weakly informative priors, despite their minimal impact, particularly for $\gamma_1$ and $\beta_1$ because they incorporate the reasonable assumption that both parameters should be positive, given the positive nature of the underlying trends (e.g. \citealt{kormendy13a}, \citealt{bosch16}).

As explained in Section \ref{sec:model and methods},
the hurdle model applied to our $244$ objects results in a fit that is not strictly linear, but rather has a linear portion and a logistic portion. Figure \ref{fig:hurdle} demonstrates how our fit deviates from the relation obtained in \cite{bosch16}. The shaded grey region represents the 95\% Bayesian credible interval and is not to be mistaken with scatter. It is a density plot of the posterior samples and does not take into account the randomness of the data generating process. This results in a more tightly constrained region than usual plots of $M_\bullet-\sigma$ scatter, such as in \cite{bosch16}. This also explains why the shaded region does not necessarily capture the data, especially for low velocity dispersion values. The solid black line represents the expected value or the mean of this posterior distribution. The implications of our hurdle model for the low-mass end, transition region and high-mass end will be discussed further in Sections \ref{subsubsec:lowmass}, \ref{subsubsec:transition} and \ref{subsubsec:highmass}, respectively.

Figure \ref{fig:compareupplims} highlights the importance of accounting for upper limits in the analysis. It compares the fit depicted in Figure \ref{fig:hurdle} with an alternative fit, in green, where upper limits are treated in the same manner as the precise measurements. The use of a mixture model in the former case drags the curve downwards, particularly at lower velocity dispersions where upper limits are more prevalent. This is the result of our model accounting for the possibility that these upper limits represent undermassive BHs or simply nondetections, thus lowering the mean BH mass for small galaxies. Consequently, the linear portion follows $M_\bullet \propto \sigma^{5.8}$, exhibiting a steeper trend compared to findings in \cite{ferrarese00}, \cite{gultekin09a} and \cite{bosch16}. The comparison with the analysis of \cite{bosch16} is especially significant as we used a nearly identical data sample. This steepening in the linear portion is a result of our approach to incorporating the upper limits. Thus, it heavily relies on the assumption that the upper limits' mass distribution is well represented by our mixture model.

The variability in velocity dispersion measurements was argued in \cite{tremaine02} to be an important consideration when analyzing the $M_\bullet-\sigma$ relation. In our study, we prioritize $\sigma_e$ over $\sigma_c$, as $\sigma_e$
provides a more accurate representation of the three-dimensional velocity structure within the half-light radius (e.g. \citealt{gultekin09a}, \citealt{bosch16}). This choice is supported by the more recent findings of \cite{gultekin09a}, who argue that there is no systematic bias when comparing $\sigma_e$
and $\sigma_c$. Specifically, $\sigma_e$ is defined by integrating the contributions of both the velocity dispersion and the rotational component, weighted by the galaxy's surface brightness (e.g. \citealt{falcon-barroso_stellar_2017}). This helps to mitigate potential uncertainties arising from differences in velocity dispersion measurements across studies.

Using the logistic portion, we find that the $50$\%, $90$\% and $99$\% probability of hosting a central BH is attained at $\sigma$ values greater than $11$, $34$ and $126$ km/s, corresponding to predicted BH masses of $11.1$, $9.8 \times 10^3$ and $1.7 \times 10^7 M_\odot$, respectively. Interestingly, a velocity dispersion of $11$ km/s is even smaller than that of the Small Magellanic Cloud (SMC), a nearby dwarf galaxy (e.g. \citealt{harris06}, \citealt{diteodoro19}). At $34$ km/s, we reach a velocity dispersion on the order of the Large Magellanic Cloud (LMC), another dwarf companion to our Milky Way, whereas a velocity dispersion of $126$ km/s approaches that of the Milky Way itself (e.g. \citealt{bekki05}, \citealt{valenti18}). While the SMC has little evidence pertaining to a central BH (e.g. \citealt{lazzarini19}), \cite{boyce17} explore the possibility that a galaxy as small as the LMC contains a BH. 

As seen in Table \ref{tab:coefficients}, the logistic parameters $\beta_0$
 and $\beta_1$ are weakly constrained, meaning the values stated in the previous paragraph are highly uncertain. This is emphasized by the wide posterior distributions of $\beta_0$ and $\beta_1$ in Figure \ref{fig:param post dist}. The nature of our dataset, which has not been corrected for selection effects, further amplifies this uncertainty (e.g. \citealt{shen_catalog_2011}).  Not accounting for selection effects can skew our understanding of the BH occupation fraction, particularly with a heterogeneous dataset like ours (e.g. \citealt{reines15}, \citealt{miller_x-ray_2015}). This bias arises because our sample is neither volume-limited nor unbiased, leading to an overrepresentation of certain types of galaxies, especially those with more easily detectable supermassive BHs. For example, many of the galaxies in our dataset are AGNs because they are more readily detectable due to their high luminosity and distinctive emission features (e.g. \citealt{kauffmann_host_2003}, \citealt{heckman14}). This preference for detecting AGNs contributes to the bias, as AGNs are more likely to be observed compared to inactive galaxies and necessarily harbor a central BH (e.g. \citealt{genzel10}, \citealt{greene_low-mass_2012}, \citealt{kormendy13a}). Consequently, this overrepresentation may inflate estimates of the probability of hosting a central BH in certain galaxy populations (e.g. \citealt{miller_x-ray_2015}). Addressing these challenges by incorporating selection effects would necessitate a more complex approach, requiring the design of a detailed hierarchical model tailored to each type of measurement. Such an approach is beyond the scope of the current study, and therefore, we leave this for future work. Although our findings suggest that most massive galaxies, as well as several dwarf galaxies, are likely to harbor a central BH, these results must be interpreted within the context of these limitations.


 \subsubsection{Data Simulation} \label{subsubsec:postpredchecks}


 We simulated data to test our model's predictions when simulating from a broader and more complete galaxy distribution. This involved randomly selecting 1000 posterior samples and generating 1000 data points for each sample. The simulated data's stellar velocity dispersions were drawn randomly from the Balmer line velocity dispersion distribution of the DR8 Sloan Digital Sky Survey MPA-JHU Value-Added Catalog (VAC) (see \citealt{kauffmann03}, \citealt{brinchmann04}, \citealt{tremonti04}, \citealt{aihara11}). 
We employed this approach to ensure a more exact representation of the local galaxy distribution, given the constraints of our dataset, which includes only galaxies with dynamically or temporally resolved BHs. Additionally, it enables us to capture low-mass galaxies more accurately in our simulated data, a critical consideration with the advent of upcoming 30-40m telescopes allowing us to probe their central BHs (e.g. \citealt{yelda13}, \citealt{do14}). We examined both the velocity dispersion distributions from Balmer lines and forbidden lines, sourced from the VAC. Since the distributions were nearly identical, we opted to use the one from Balmer lines, given that a portion of our data sample's velocity dispersions originate from Balmer lines (e.g. \citealt{barth09}, \citealt{bennert11}, \citealt{bosch16}). Prior to drawing the velocity dispersions, we removed outliers and used a redshift cutoff of $z\approx 0.1$, which corresponds to the highest redshift galaxy in our data sample. The conclusions drawn from our study are therefore only applicable to the Universe up to this redshift, where the BH mass measurements are more reliable.

 We investigate the predictions of our model by comparing the BH mass marginal distributions of our observational data with our simulated data in Figure \ref{fig:histogram}. Our findings reveal a decent similarity between the distributions, with the simulated data (depicted in yellow) closely mirroring the observed data (depicted in blue). 
The main distinctions between the two distributions lie in their spread: the observed data shows a higher peak towards the center with a narrower spread, whereas the simulated data exhibits broader tails. These variations stem from the use of a distinct velocity dispersion distribution when generating the simulated data. Remarkably, despite drawing initial velocity dispersions for simulation from the SDSS distribution, our simulated data accurately captures the distribution of upper limits, which are prominently represented at a BH mass of zero.
 



To further assess the model's predictions, we also plot simulated data against our observational data in Figure \ref{fig:postpred}. Given the considerable uncertainty in the logistic parameters, we investigate two alternative interpretations of these simulated data. In the left panel, we consider the possibility that simulated zeros in the logistic portion represent upper limits, indicating galaxies that may lack a central BH or may simply be nondetections. In the right panel we explore the idea that these simulated zeros correspond to galaxies genuinely devoid of a BH. Thus, in the left panel, we plot the simulated zeros from the logistic portion as upper limits at their corresponding simulated mass from the linear portion, while in the right panel, they are represented at a BH mass of zero.

As discussed in Section \ref{sec:results}, galaxies with velocity dispersions less than $34$ km/s exhibit a probability of less than $10$\% of hosting a central BH. Consequently, we anticipate that galaxies lacking a central BH would predominantly fall below this $\sigma$ threshold, especially seeing as most of the galaxies from the SDSS distribution have low velocity dispersions. Contrary to this expectation, under the assumption depicted in the right panel for our simulated data, a substantial majority (approximately $75$\%) of galaxies without a central BH possess $\sigma \geq 33$ km/s. Furthermore, with the interpretation of the right panel, we find that approximately $2.1$\% of simulated data corresponds to galaxies devoid of central BHs, meanwhile they represent $0.8$\% of observed data. These inconsistencies prompt us to favor a more lenient interpretation of the simulated zeros, as proposed in the left panel, where they merely represent a potential absence of a central BH. This lenient interpretation also aligns well with our understanding that the BH occupation fraction is highly uncertain, due to selection effects and our biased data sample.



\subsubsection{Implications for the low-mass end} \label{subsubsec:lowmass}

The low-mass end of the $M_\bullet-\sigma$ relation allows us to better understand the formation of intermediate-mass BHs, which are thought to be the seeds from which supermassive BHs grow (e.g. \citealt{mezcua17}). In \cite{volonteri09} and \cite{wassenhove10}, they suggest that the low-mass end of the relationship between BH mass and velocity dispersion flattens towards an asymptotic value if BH seeds are heavy, i.e. $10^5 M_\odot$, resulting in a distinctive "plume" of ungrown BHs. They mention that this is a signature of direct collapse models (see \citealt{begelman06}). In contrast, they show that the observable scaling rules would not detect the asymptotic value if BH seeds are light, of order $10^2 M_\odot$, as anticipated by models akin to Population III stars (e.g. \citealt{alighieri08}). This is explained by the fact that the asymptotic value lies at masses below our observational limit. 


Several observational studies have found a flattening of the curve (e.g. \citealt{reines15}, \citealt{ferre-mateu15}, \citealt{mezcua17}, \citealt{martin-navarro18}), supporting direct collapse models. \cite{reines15} even observe this flattening in their relationship between BH mass and host galaxy total stellar mass. This was done using a sample of 262 broad-line active galactic nuclei and 79 galaxies with dynamically resolved BHs, excluding upper limits. In \cite{mezcua17}, they provide an overview of the observational evidence for intermediate-mass BHs and show that they also observe this "plume", around $10^5-10^6 M_\odot$ when plotting them against the $M_\bullet-\sigma$ relation derived in \cite{kormendy13a}. They mention that this could possibly be due to observational biases, which are further discussed in \cite{pacucci18}. \cite{pacucci18} simulate the coevolution of central BHs and their host galaxies using a semi-analytical model. Their results predict a deviation from the $M_\bullet-\sigma$ relation around $10^5 M_\odot$, resulting in under-massive BHs. They suggest that this could be caused by the fact that smaller BHs exhibit inefficient growth and are incapable of producing substantial outflows sufficient to initiate the growth-regulation mechanism. 

In a more recent study, \cite{greene20} find no evidence for a change in the $M_\bullet-\sigma$ relation at low $\sigma$ values, but rather a higher scatter. This results in a wider distribution of both over-massive and under-massive BHs when considering the low-mass end. 

Our model predicts under-massive BHs, ranging from $\approx 10-10^5 M_\odot$, when compared to the scaling relation from \cite{bosch16} for the low-mass end of the fit. This is similar to the result from the simulations in \cite{pacucci18} and further supports the theory that the flattening in the $M_\bullet-\sigma$ relation is simply due to our limited observational capacities. This also suggests a higher number of intermediate-mass BHs and BH seeds formed by the collapse of Population III stars or via mergers in dense stellar clusters (e.g. \citealt{mezcua17}). Although our results are in discordance with direct collapse models for the origin of seed BHs, this could be biased by our assumption that our model for the upper limits is realistic. If the true masses of these BHs are significantly closer to the upper limit, this assumption may force the low-mass end of our fit to be pulled excessively downward.

Identifying BHs in lower mass galaxies, such as the SMC and the LMC, presents unique challenges that may contribute to the observed discrepancies in the mass estimates and their relationships with host galaxy parameters (e.g. \citealt{lazzarini19}). The irregular morphology and complex stellar dynamics of these galaxies complicate the determination of the galactic center, which is crucial for accurate measurements of key parameters needed for BH identification (e.g. \citealt{diteodoro19}). Furthermore, the low luminosity of these systems can hinder detection, particularly when using traditional methods that rely on observing the motion of surrounding stars (e.g. \citealt{boyce17}). These difficulties emphasize the need for a careful interpretation of our findings regarding the low-end of the $M_\bullet-\sigma$ relation.

The existence of additional under-massive BHs bodes well for the upcoming generation of giant segmented-mirror telescopes, such as the Extremely Large Telescope and the Thirty Meter Telescope. These telescopes will achieve unprecedented infrared resolution and sensitivity (e.g., \citealt{yelda13}). Their observational capacities will enable the definitive detection of galaxies hosting intermediate-mass BHs, approximately $\sim 10^4 M_\odot$, within the local group (e.g., \citealt{do14}). This corresponds to galaxies with velocity dispersions of $35$ km/s given the linear portion and we expect $90$\% of them to host a central BH if our hurdle model is correct. As indicated by our simulated data in Figures \ref{fig:histogram} and \ref{fig:postpred}, the number of potential candidates in this mass range is likely more extensive than previously envisioned.

\subsubsection{Implications for the transition region} \label{subsubsec:transition}

\cite{martin-navarro18} derive the $M_\bullet-\sigma$ relation from a sample of only low-mass galaxies. This results in a fit that overestimates the relation from \cite{bosch16} until a transition mass of $3.4$ $\pm$ $2.1$ x $10^{10} M_\odot$, after which it underestimates it. Conversely, our model predicts under-massive BHs until a transition mass of $\approx 1.8$ x $10^7 M_\odot$, whereafter it predicts over-massive BHs. This discrepancy can be explained by several factors. The linear $M_\bullet-\sigma$ relation is known to flatten at the low-mass end and \cite{martin-navarro18} use a data sample comprised of only low-mass galaxies, resulting in a flattened curve. Meanwhile, our data sample contains both low-mass and high-mass galaxies. In addition, our model takes into consideration the upper limits by treating them with a mixture model, which also tends to drag the curve downward.

\subsubsection{Implications for the high-mass end} \label{subsubsec:highmass}

When considering the high-mass end, there is a discrepancy between the $M_\bullet-\sigma$ relation and the $M_\bullet-L$ relation (e.g. \citealt{lauer07}, \citealt{mezcua18a}, \citealt{dullo21}). The latter is an empirical correlation between central BH mass and its host galaxy's V-band/B-band/K-band/R-band luminosity (see \citealt{faber89}, \citealt{faber97}, \citealt{merritt01b}, \citealt{graham07}). This inconsistency is thought to be due to the presence of brighest cluster galaxies (BCGs), whose stellar velocity dispersion increases weakly with luminosity although they harbor the most massive BHs (e.g. \citealt{oegerle91}, \citealt{boylan-kolchin06}, \citealt{mcnamara09}, \citealt{hlavacek-larrondo12}, \citealt{mezcua18a}). 

In \cite{lauer07}, they mention that the $M_\bullet-L$ relation predicts masses near $10^{10} M_\odot$ for the most luminous BCGs, whereas the $M_\bullet-\sigma$ relation predicts masses inferior to $3.0$ x $10^9 M_\odot$. This is also observed in \cite{hlavacek-larrondo12} and \cite{mezcua18a}. In the latter study they find that $\approx 40\%$ of the 72 BCGs in their sample should host ultramassive BHs ($M_\bullet > 10^{10} M_\odot$) if they are to be found on the FP of BH accretion (e.g. \citealt{merloni03}, \citealt{bosch16}, \citealt{gultekin19}). This is of course not supported by the linear scaling relation involving stellar velocity dispersion, which underestimates these BH masses when compared to the FP detailed in \cite{mezcua18a}.

\cite{dullo21} find that BH masses at the center of large-core galaxies, which make up the majority of BCGs (e.g. \citealt{laine03}), are underestimated by $2.5-4 \sigma_s$ when predicted by the $M_\bullet-\sigma$ relation. This suggests a steepening in this linear relation when considering the high-mass end. Our results agree with this prediction when employing the hurdle model. As demonstrated in Figure \ref{fig:hurdle}, our fit predicts over-massive BHs for high values of $\sigma$ ($\gtrapprox 127$ km/s) when comparing to the relation derived in \cite{bosch16}. Our hurdle model also predicts more BHs with masses greater than $10^{10} M_\odot$ starting at velocity dispersion values of $\approx 381$ km/s. Indeed, although these BHs only represent $0.8$\% of our observed data, they represent $6.3$\% of our simulated data.

The presence of ultramassive BHs at the high-mass end has several implications pertaining to our understanding of BH feedback in these host galaxies, which are most likely BCGs (e.g. \citealt{collins_early_2009}). It also has implications for the FP of BH accretion. As of now, most observational evidence shows that these BHs are positively offset from the FP of BH activity (e.g. \citealt{dalla_bonta06}, \citealt{mcconnell12}, \citealt{ishibashi17}, \citealt{mezcua18a}, \citealt{dullo21}). \citealt{mezcua18a} mention that BCGs should sit on average on the FP if they are governed by the same accretion physics as lower mass galaxies. \cite{mcconnell12} mentions that these systematic offsets allude to distinctive evolutionary mechanisms for galaxies close to cluster cores. Thus, \cite{mezcua18a} argue against the notion of estimating BH masses in BCGs using FP correlations that favour the jet model (such as in \citealt{kording06}, \citealt{plotkin12}). This is supported by \cite{dullo21}, in which they mention that such models must treat extremely massive galaxies independently. Our findings also support the existence of a $M_\bullet - V_c$ correlation, where $V_c$ is the host galaxy's circular velocity, a proxy for dark matter halo mass. This correlation has been under debate for a long time (e.g. \citealt{volonteri11}, \citealt{bansal22}), but is supported by the presence of ultramassive BHs in BCGs (e.g. \citealt{mezcua18a}).

Furthermore, the existence of more ultramassive BHs is advantageous for the Event Horizon Telescope (EHT) as these present more candidates for imaging. Ultramassive BHs may be ideal candidates for the EHT as they have greater angular diameters (e.g. EHT \citealt{eht19-1}). In addition, they tend to have lower angular momentum, resulting in easier targets as matter rotates the BH at lower velocities (e.g. EHT \citealt{eht19-1}, \citealt{sisk-reynes22}).  These images are crucial as they deepen and confirm our understanding of the underlying physics of BHs (see EHT \citealt{eht19-1, eht19-2, eht19-3, eht19-4, eht19-5, eht19-6, eht21-7, eht21-8, eht22-1, eht22-2, eht22-3, eht22-4, eht22-5, eht22-6}). 

The two supermassive BHs imaged by the EHT are M87* and Sagittarius A*, which are central BHs to the host galaxies M87 and the Milky Way, respectively (e.g. EHT \citealt{eht19-1}, \citealt{eht22-1}). They are located at distances of $16.8$ Mpc and $0.0083$ Mpc while possessing respective masses of $6.2\times10^9 M_\odot$ and $4.1\times10^6 M_\odot$ (e.g. EHT \citealt{eht19-2}). This results in angular diameters of $\sim 40$ $\mu as$ and $\sim 50$ $\mu as$, respectively (e.g. EHT \citealt{eht19-4}, \citealt{eht22-1}). As an example, an ultramassive BH with $M_\bullet = 10^{10} M_\odot$, which is reasonable given the abundance of these ultramassive BHs in our simulated data of Figure \ref{fig:postpred}, could have an angular diameter greater than $\sim 40$ $\mu as$ up to a distance of $\sim 25$ Mpc resulting in a redshift of $z\sim 0.0059$ (e.g. EHT \citealt{eht19-5}). BHs with masses of $M_\bullet = 5 \times 
 10^{10} M_\odot$ and $M_\bullet = 9 \times 10^{10} M_\odot$ yield the following results: $\sim 128$ Mpc and $z\sim 0.0305$, as well as $\sim 230$ Mpc and $z\sim 0.0543$, respectively. Thus, our findings suggest that the number of candidates for imaging by the EHT could increase significantly as ultramassive BHs up to a redshift of $z\sim 0.0543$ may be considered.

\section{Conclusion} \label{sec:conclusion}

In this paper, we apply a hurdle model to the $M_\bullet-\sigma$ relation. We use a sample of 244 galaxies for which the mass of the central BH has been precisely resolved or has been constrained to an upper limit. This hierarchical model allowed us to account for the probability that a galaxy will contain a central BH while simultaneously accounting for the relation between $M_\bullet$ and $\sigma$. We obtain the following values for the hurdle model parameters (see Section \ref{subsec:hurdle model}): $\beta_0=-4.4$, $\beta_1=4.3$, $\gamma_0=-4.9$, $\gamma_1=5.8$ and $\lambda=0.98$, with 95\% Bayesian credible intervals of ($-8.9, 0.2$), ($1.9, 6.9$), ($-5.3, -4.4$), ($5.6, 6.0$) and ($0.92, 1.00$), respectively.

Our findings suggest that $M_\bullet \propto \sigma^{5.8}$, by using the linear portion of our curve. This result is steeper than most studies (e.g. \citealt{ferrarese00}, \citealt{gultekin09a}, \citealt{bosch16}) and has significant effects on our understanding of galaxy and BH coevolution. It can be explained by our method of including upper limits in the analysis, as the assumption that the upper limits are well represented by our mixture model tends to steepen the curve.

Using the logistic portion of our hurdle model, we found that galaxies with $\sigma$ values greater than $11$, $34$ and $126$ km/s have a 50\%, 90\% and 99\% probability of hosting a central BH. Thus, we suggest that most massive galaxies as well as many dwarf galaxies likely harbor a central BH. Although it should be noted that the previous values are highly uncertain due to the weakly constrained logistic parameters. 


When comparing the hurdle model to linear models from other studies, we find that it predicts under-massive BHs, from $10-10^5$ $M_\odot$, when considering galaxies at the low-mass end. These results are similar to those of models related to 
Population III star seeds and further support the fact that the flattening in the $M_\bullet-\sigma$ is due to our limited observational capacities, as discussed in Section \ref{subsubsec:lowmass} (e.g. \citealt{alighieri08}, \citealt{mezcua17}, \citealt{pacucci18}). Though this could be biased by our assumptions on the mass distribution of the upper limits. 

We also find that our fit predicts over-massive BHs when considering galaxies with high stellar velocity dispersion (i.e. $\sigma \gtrapprox 127$ km/s). This result implies the presence of more ultramassive BHs which we suggest are most likely at the center of BCGs. Hence, our results are in favor of the existence of a $M_\bullet-V_c$ correlation (e.g. \citealt{mezcua18a}). We also mention that this has implications for our understanding of BH feedback in these galaxies, as they should sit on the FP of BH accretion if they are governed by the same accretion physics as lower $\sigma$ galaxies (e.g. \citealt{mezcua18a}). Our model predicts them to be positively offset, suggesting that this is not the case (e.g. \citealt{mcconnell12}). 

The transition between our model predicting under-massive BHs to predicting over-massive BHs happens at $M_\bullet \approx 1.8 \times 10^7 M_\odot$. This is lower than the transition mass in other studies, such as \cite{martin-navarro18}, but can be explained by the fact that we include the upper limits with a two part model, in which the logistic portion tends to drag the curve downward. 

Improvements could be made to our analysis by employing more informative priors on the hurdle model parameters, as opposed to using the weak priors from Section \ref{subsec:hurdle model}. 
Although this would require extensive analysis of the galaxies with upper limits, it holds significance for the parameters in the logistic portion due to their large uncertainties. We may also consider defining a probability distribution that more accurately reflects the mass distribution of upper limits as constraints tighten, rather than relying on the assumption of our mixture model. Another improvement could involve incorporating selection effects into our Bayesian framework, following a similar approach as demonstrated in \cite{leistedt16}. This addition could specifically target the upper limits within our dataset, distinguishing galaxies for which we are highly confident in the absence of evidence related to a central BH from those where there is only a marginal probability. The integration of such considerations into the model has the potential to better constrain the logistic portion, while also rendering this aspect of the two-step process more interpretable and less nuanced when treating upper limits.

In our study, we sought to better understand the M-sigma relation. However, when utilizing scaling relations to estimate BH masses, it is essential to consider alternative approaches that may yield more robust insights. Notably, the literature suggests that the stellar mass-BH mass relation serves as a reliable proxy across various galaxy types (e.g. \citealt{kormendy13a}, \citealt{reines15}, \citealt{greene20}). While the M-sigma relation has its merits, especially in dynamically dominated systems, the correlation between stellar mass and BH mass has been well established within the context of empirical scaling relations (e.g. \citealt{ferrarese00}). This presents an opportunity for future studies to explore the implications of utilizing stellar mass in conjunction with our findings regarding the M-sigma relation.

\section*{Acknowledgments}

J.H.-L. acknowledges support from NSERC via the Discovery grant program, as well as the Canada Research Chair program. M. M. acknowledges support from the Spanish Ministry of Science and Innovation through the project PID2021-124243NB-C22. This work was partially supported by the program Unidad de Excelencia Maria de Maeztu CEX2020-001058-M. G.M.E. acknowledges support from NSERC via the Discovery grant program (RGPIN-2020-04554).

\section*{Data Availability} \label{sec:cite}

All methods and data used in this paper are available at \href{https://github.com/GabrielSasseville01/BH_M-sigma_compilation}{\faGithubSquare  GabrielSasseville01}.
\appendix
\restartappendixnumbering

\section{Posterior Distributions}
\label{appendix:posterior}

\begin{figure*}[ht]
\plotone{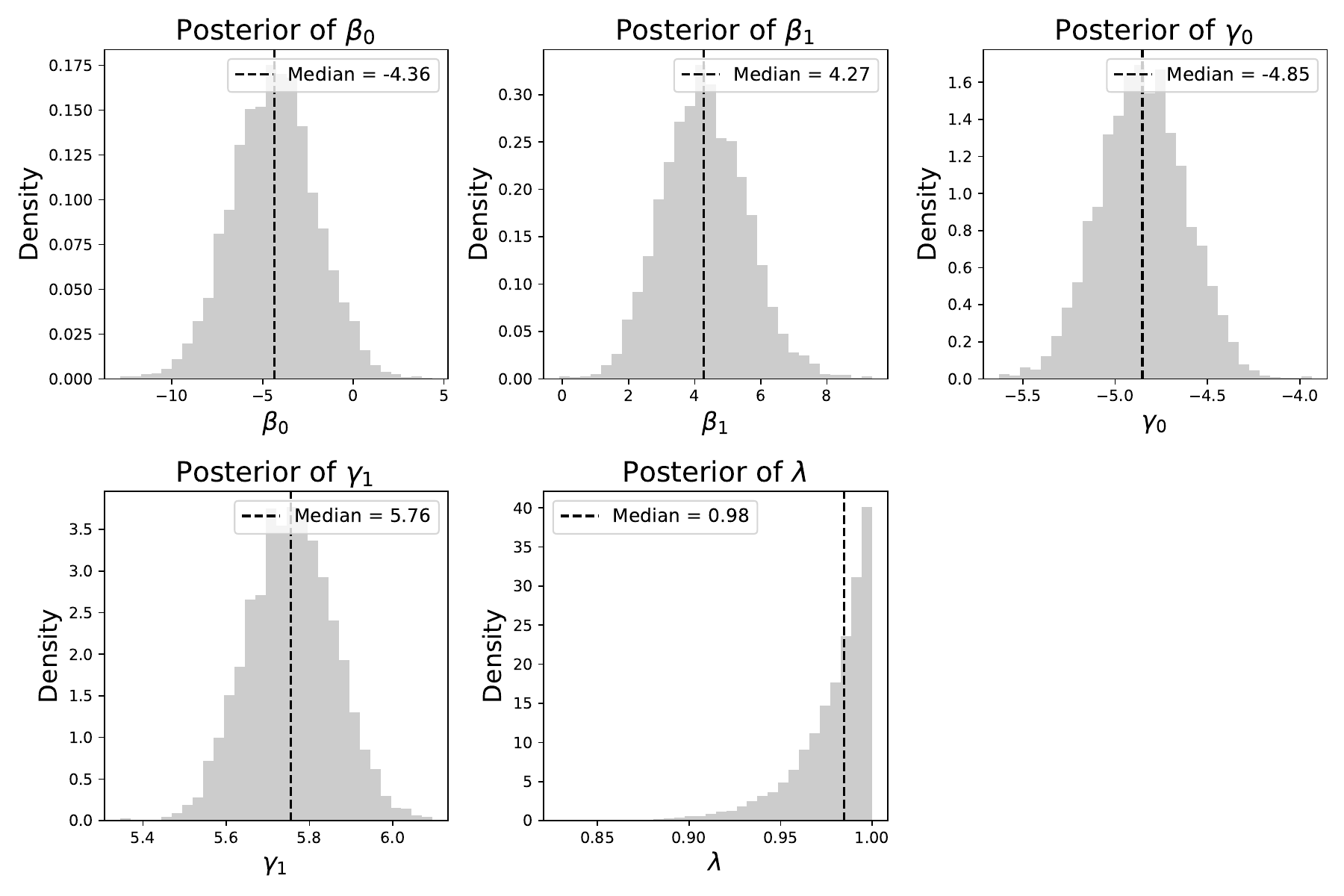}
\epsscale{0.2}
\caption{Posterior distributions of the hurdle parameters: $\beta_0$, $\beta_1$, $\gamma_0$, $\gamma_1$ and $\lambda$. The black dashed lines represent their respective median values.}
\label{fig:param post dist}
\end{figure*}

\begin{figure}[ht]
\plotone{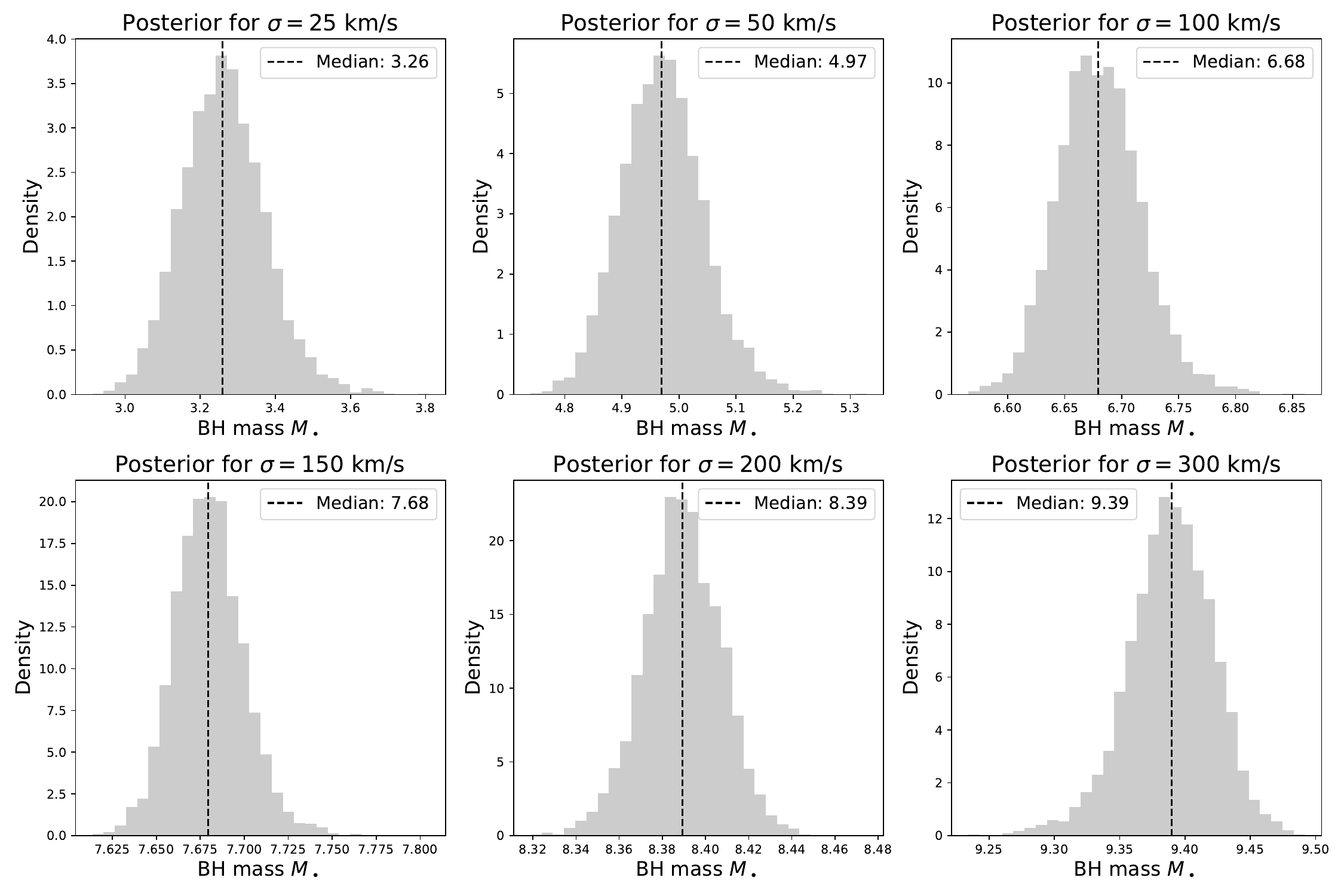}
\epsscale{0.1}
\caption{Black hole mass posterior distributions conditioned on several values of velocity dispersion: $\sigma=25$, $\sigma=50$, $\sigma=100$, $\sigma=150$, $\sigma=200$ and $\sigma=300$ km/s. The black dashed lines represent their respective median values.}
\label{fig:bh mass post dist}
\end{figure}

In this appendix, we present posterior distributions of interest. The posterior distributions over all hurdle model parameters are shown in Figure \ref{fig:param post dist}. Figure \ref{fig:bh mass post dist} shows the BH mass posterior distributions conditioned on several values of velocity dispersion, notably $\sigma=25$, $\sigma=50$, $\sigma=100$, $\sigma=150$, $\sigma=200$ and $\sigma=300$ km/s.

\section{Data}

In this appendix, we present our data compilation. Table \ref{tab:included} consists of the data included in the fits, while Table \ref{table:excluded} consists of the data excluded from the fits.

\startlongtable
\begin{longrotatetable}
\begin{deluxetable*}{lrrrhrcl}
\tablecaption{Sample of the Black Hole Mass Measurements Included in the Fits\label{tab:included}}
\tablewidth{700pt}
\tabletypesize{\scriptsize}
\tablehead{
\colhead{Name} & \colhead{Distance} & 
\colhead{$M_\bullet$} & \colhead{$\sigma_e$} & \colhead{} &
\colhead{Method} & \colhead{$M_\bullet$ Uncertainty} & \colhead{References}   \\ 
\colhead{} & \colhead{(Mpc)} & \colhead{log($M_\odot$)} & \colhead{log(km $s^{-1}$)} & 
\colhead{} & \colhead{} & \colhead{$1/2/3\sigma_s$} & \colhead{} 
} 
\decimalcolnumbers
\startdata
3C120 & 141.4$\pm$14.1 & 7.73$\pm$0.46 & 2.21$\pm$0.05 & 0.03$\pm$0.00 & reverb & 1&\cite{kollatschny14} \\ 
3C390.3 & 240.3$\pm$24.0 & 8.62$\pm$0.47 & 2.44$\pm$0.03 & nan$\pm$nan & reverb & 1&\cite{dietrich12} \\ 
A1836BCG & 152.4$\pm$8.4 & 9.57$\pm$0.18 & 2.46$\pm$0.02 & 0.04$\pm$0.00 & gas & 1&\cite{bonta09} \\ 
A3565BCG & 49.2$\pm$3.6 & 9.11$\pm$0.22 & 2.51$\pm$0.02 & 0.01$\pm$0.00 & gas & 1&\cite{bonta09} \\ 
Ark120 & 140.1$\pm$14.0 & 8.05$\pm$0.51 & 2.28$\pm$0.02 & 0.03$\pm$0.00 & reverb & 1&\cite{doroshenko08} \\ 
Arp151 & 90.3$\pm$9.0 & 6.65$\pm$0.49 & 2.07$\pm$0.01 & 0.02$\pm$0.00 & reverb & 1&\cite{bentz10} \\ 
Circinus & 2.8$\pm$0.5 & 6.06$\pm$0.31 & 2.20$\pm$0.05 & 0.00$\pm$0.00 & maser & 1&\cite{greenhill03} \\ 
CygnusA & 257.1$\pm$25.7 & 9.41$\pm$0.38 & 2.43$\pm$0.02 & 0.06$\pm$0.00 & gas & 1&\cite{tadhunter03} \\ 
ESO558-009 & 102.5$\pm$10.2 & 7.22$\pm$0.09 & 2.23$\pm$0.05 & 0.03$\pm$0.00 & maser & 1&\cite{greene16} \\ 
Fairall9 & 201.4$\pm$20.1 & 8.27$\pm$0.63 & 2.35$\pm$0.08 & 0.05$\pm$0.00 & reverb & 1&\cite{peterson04,vdb15} \\ 
Henize2-10 & 9.0$\pm$0.9 & 0.00$^{+7.00}_{-0.00}$ & 1.54$\pm$0.19 & 0.00$\pm$0.00 & star & 3&\cite{nguyen14} \\ 
IC0342 & 3.7$\pm$0.4 & 6.41$^{+0.88}_{-6.41}$ & 1.78$\pm$0.17 & 0.00$\pm$0.00 & gas & 1&\cite{beifiori12} \\ 
IC1459 & 28.9$\pm$3.7 & 9.39$\pm$0.24 & 2.53$\pm$0.02 & 0.01$\pm$0.00 & star & 1&\cite{cappellari02} \\ 
IC1481 & 89.9$\pm$9.0 & 7.15$\pm$0.40 & 1.98$\pm$0.12 & 0.02$\pm$0.00 & maser & 1&\cite{hure11,vdb15} \\ 
IC2560 & 41.8$\pm$4.2 & 7.64$\pm$0.05 & 2.15$\pm$0.03 & 0.01$\pm$0.00 & maser & 3&\cite{greene16} \\ 
IC3639 & 44.8$\pm$4.5 & 7.01$^{+1.06}_{-7.01}$ & 1.96$\pm$0.02 & 0.01$\pm$0.00 & gas & 1&\cite{beifiori12} \\ 
J0437+2456 & 66.0$\pm$6.6 & 6.45$\pm$0.09 & 2.04$\pm$0.05 & nan$\pm$nan & maser & 1&\cite{greene16} \\ 
Mrk0050 & 100.3$\pm$10.0 & 7.40$\pm$0.53 & 2.03$\pm$0.06 & 0.02$\pm$0.00 & reverb & 1&\cite{barth11a} \\ 
Mrk0079 & 95.0$\pm$9.5 & 7.58$\pm$0.70 & 2.11$\pm$0.04 & 0.02$\pm$0.00 & reverb & 1&\cite{peterson04} \\ 
Mrk0202 & 90.0$\pm$9.0 & 6.11$\pm$0.85 & 1.89$\pm$0.02 & 0.02$\pm$0.00 & reverb & 1&\cite{bentz10} \\ 
Mrk0509 & 147.3$\pm$14.7 & 8.03$\pm$0.44 & 2.26$\pm$0.03 & 0.03$\pm$0.00 & reverb & 1&\cite{peterson04} \\ 
Mrk0590 & 113.0$\pm$11.3 & 7.55$\pm$0.54 & 2.28$\pm$0.01 & 0.03$\pm$0.00 & reverb & 1&\cite{peterson04} \\ 
Mrk0817 & 134.7$\pm$13.5 & 7.57$\pm$0.54 & 2.08$\pm$0.05 & 0.03$\pm$0.00 & reverb & 1&\cite{denney10} \\ 
Mrk1216 & 94.0$\pm$9.4 & 9.30$^{+2.10}_{-9.30}$ & 2.49$\pm$0.01 & 0.02$\pm$0.00 & star & 1&\cite{yildirim15} \\ 
Mrk1310 & 83.8$\pm$8.4 & 6.19$\pm$0.58 & 1.92$\pm$0.03 & 0.02$\pm$0.00 & reverb & 1&\cite{bentz10} \\ 
NGC0193 & 49.7$\pm$5.0 & 8.40$\pm$0.96 & 2.27$\pm$0.04 & 0.01$\pm$0.00 & gas & 1&\cite{beifiori12} \\ 
NGC0205 & 0.7$\pm$0.1 & 0.00$^{+4.34}_{-0.00}$ & 1.59$\pm$0.07 & -0.00$\pm$0.00 & star & 3&\cite{valluri05} \\ 
NGC0221 & 0.8$\pm$0.0 & 6.39$\pm$0.58 & 1.87$\pm$0.02 & -0.00$\pm$0.00 & star & 1&\cite{vdb10} \\ 
NGC0289 & 17.1$\pm$1.7 & 7.38$^{+0.89}_{-7.38}$ & 2.03$\pm$0.05 & 0.01$\pm$0.00 & gas & 1&\cite{beifiori12} \\ 
NGC0307 & 52.8$\pm$5.7 & 8.60$\pm$0.18 & 2.31$\pm$0.01 & 0.01$\pm$0.00 & star & 1&\cite{saglia16} \\ 
NGC0315 & 57.7$\pm$5.8 & 8.92$\pm$0.94 & 2.49$\pm$0.04 & 0.02$\pm$0.00 & gas & 1&\cite{beifiori12} \\ 
NGC0383 & 59.2$\pm$5.9 & 8.76$\pm$0.97 & 2.38$\pm$0.03 & 0.02$\pm$0.00 & gas & 1&\cite{beifiori12} \\ 
NGC0404 & 3.1$\pm$0.3 & 5.65$\pm$0.76 & 1.39$\pm$0.09 & -0.00$\pm$0.00 & gas & 1&\cite{seth10} \\ 
NGC0428 & 16.1$\pm$1.6 & 4.48$^{+0.37}_{-4.48}$ & 1.50$\pm$0.07 & 0.00$\pm$0.00 & star & 3&\cite{neumayer12} \\ 
NGC0524 & 24.2$\pm$2.2 & 8.94$\pm$0.16 & 2.37$\pm$0.02 & 0.01$\pm$0.00 & star & 1&\cite{krajnovic09} \\ 
NGC0541 & 63.7$\pm$6.4 & 8.59$\pm$1.02 & 2.28$\pm$0.01 & 0.02$\pm$0.00 & gas & 1&\cite{beifiori12} \\ 
NGC0584 & 22.2$\pm$1.6 & 8.15$\pm$0.57 & 2.28$\pm$0.70 & 0.01$\pm$0.00 & star & 1&\cite{thater19} \\ 
NGC0598 & 0.8$\pm$0.1 & 0.00$^{+9.44}_{-0.00}$ & 1.26$\pm$0.20 & -0.00$\pm$0.00 & star & 1&\cite{gebhardt01,merritt01b} \\ 
NGC0613 & 15.4$\pm$1.5 & 7.60$\pm$1.05 & 2.09$\pm$0.07 & 0.00$\pm$0.00 & gas & 1&\cite{beifiori12} \\ 
NGC0741 & 65.7$\pm$6.6 & 8.67$\pm$1.11 & 2.37$\pm$0.02 & 0.02$\pm$0.00 & gas & 1&\cite{beifiori12} \\ 
NGC0788 & 47.9$\pm$4.8 & 7.92$\pm$0.84 & 2.10$\pm$0.06 & 0.01$\pm$0.00 & gas & 1&\cite{beifiori12} \\ 
NGC0821 & 23.4$\pm$1.8 & 8.22$\pm$0.63 & 2.32$\pm$0.02 & 0.01$\pm$0.00 & star & 1&\cite{schulze11} \\ 
NGC1023 & 10.8$\pm$0.8 & 7.62$\pm$0.17 & 2.22$\pm$0.02 & 0.00$\pm$0.00 & star & 1&\cite{bower01} \\ 
NGC1042 & 18.2$\pm$1.8 & 4.40$^{+2.08}_{-4.40}$ & 1.64$\pm$0.07 & 0.00$\pm$0.00 & star & 3&\cite{neumayer12} \\ 
NGC1052 & 18.1$\pm$1.8 & 8.24$\pm$0.88 & 2.28$\pm$0.01 & 0.00$\pm$0.00 & gas & 1&\cite{beifiori12} \\ 
NGC1097 & 14.5$\pm$1.5 & 8.14$\pm$0.27 & 2.29$\pm$0.01 & 0.00$\pm$0.00 & CO & 1&\cite{onishi15} \\ 
NGC1194 & 58.0$\pm$6.3 & 7.85$\pm$0.15 & 2.17$\pm$0.07 & 0.01$\pm$0.00 & maser & 1&\cite{kuo11} \\ 
NGC1271 & 84.0$\pm$8.4 & 9.53$\pm$0.17 & 2.45$\pm$0.01 & 0.02$\pm$0.00 & star & 1&\cite{walsh15} \\ 
NGC1275 & 70.0$\pm$7.0 & 8.98$\pm$0.60 & 2.39$\pm$0.05 & 0.02$\pm$0.00 & gas & 1&\cite{scharwachter13,vdb15} \\ 
NGC1277 & 71.0$\pm$7.1 & 9.67$\pm$0.44 & 2.50$\pm$0.01 & 0.02$\pm$0.00 & star & 1&\cite{walsh16} \\ 
NGC1316 & 18.6$\pm$0.6 & 8.18$\pm$0.76 & 2.35$\pm$0.02 & 0.01$\pm$0.00 & star & 1&\cite{nowak08} \\ 
NGC1320 & 49.1$\pm$4.9 & 6.74$\pm$0.48 & 2.15$\pm$0.05 & 0.01$\pm$0.00 & maser & 1&\cite{greene16} \\ 
NGC1332 & 22.3$\pm$1.9 & 8.83$\pm$0.13 & 2.52$\pm$0.02 & 0.01$\pm$0.00 & CO & 1&\cite{barth16b} \\ 
NGC1358 & 48.2$\pm$4.8 & 8.37$\pm$0.97 & 2.23$\pm$0.05 & 0.01$\pm$0.00 & gas & 1&\cite{beifiori12} \\ 
NGC1374 & 19.2$\pm$0.7 & 8.76$\pm$0.19 & 2.31$\pm$0.02 & 0.00$\pm$0.00 & star & 1&\cite{rusli13} \\ 
NGC1386 & 16.1$\pm$1.6 & 6.07$\pm$0.88 & 2.22$\pm$0.02 & 0.00$\pm$0.00 & maser & 1&\cite{braatz97} \\ 
NGC1398 & 24.8$\pm$4.1 & 8.03$\pm$0.25 & 2.37$\pm$0.01 & 0.00$\pm$0.00 & star & 1&\cite{saglia16} \\ 
NGC1399 & 20.9$\pm$0.7 & 8.94$\pm$0.92 & 2.53$\pm$0.02 & 0.00$\pm$0.00 & star & 1&\cite{gebhardt07,houghton06} \\ 
NGC1407 & 28.0$\pm$3.4 & 9.65$\pm$0.24 & 2.45$\pm$0.02 & 0.01$\pm$0.00 & star & 1&\cite{rusli13} \\ 
NGC1428 & 20.7$\pm$2.1 & 0.00$^{+7.00}_{-0.00}$ & 1.91$\pm$0.02 & 0.01$\pm$0.00 & star & 3&\cite{lyubenova13} \\ 
NGC1493 & 11.4$\pm$1.1 & 5.40$^{+0.51}_{-5.40}$ & 1.55$\pm$0.07 & 0.00$\pm$0.00 & star & 3&\cite{neumayer12} \\ 
NGC1497 & 75.3$\pm$7.5 & 8.63$\pm$0.56 & 2.39$\pm$0.04 & 0.02$\pm$0.00 & gas & 1&\cite{beifiori12} \\ 
NGC1550 & 51.6$\pm$5.6 & 9.57$\pm$0.07 & 2.48$\pm$0.02 & 0.01$\pm$0.00 & star & 3&\cite{rusli13} \\ 
NGC1600 & 64.0$\pm$6.4 & 10.23$\pm$0.12 & 2.47$\pm$0.02 & 0.02$\pm$0.00 & star & 1&\cite{thomas16} \\ 
NGC1667 & 56.1$\pm$5.6 & 8.20$\pm$0.69 & 2.24$\pm$0.07 & 0.02$\pm$0.00 & gas & 1&\cite{beifiori12} \\ 
NGC1961 & 48.6$\pm$4.9 & 8.29$\pm$1.03 & 2.34$\pm$0.08 & 0.01$\pm$0.00 & gas & 1&\cite{beifiori12} \\ 
NGC2110 & 29.1$\pm$2.9 & 8.12$\pm$1.93 & 2.30$\pm$0.05 & 0.01$\pm$0.00 & gas & 1&\cite{beifiori12} \\ 
NGC2179 & 35.8$\pm$3.6 & 8.31$\pm$0.70 & 2.19$\pm$0.03 & 0.01$\pm$0.00 & gas & 1&\cite{beifiori12} \\ 
NGC2273 & 29.5$\pm$1.9 & 6.93$\pm$0.11 & 2.16$\pm$0.05 & 0.01$\pm$0.00 & maser & 1&\cite{kuo11} \\ 
NGC2329 & 72.3$\pm$7.2 & 8.18$\pm$0.54 & 2.34$\pm$0.03 & 0.02$\pm$0.00 & gas & 1&\cite{beifiori12} \\ 
NGC2549 & 12.7$\pm$1.6 & 7.16$\pm$1.10 & 2.15$\pm$0.02 & 0.00$\pm$0.00 & star & 1&\cite{krajnovic09} \\ 
NGC2685 & 12.5$\pm$1.3 & 6.59$^{+1.24}_{-6.59}$ & 2.02$\pm$0.02 & 0.00$\pm$0.00 & gas & 1&\cite{beifiori12} \\ 
NGC2748 & 23.4$\pm$8.2 & 7.65$\pm$0.72 & 2.06$\pm$0.02 & 0.00$\pm$0.00 & gas & 1&\cite{atkinson05} \\ 
NGC2778 & 23.4$\pm$2.3 & 7.16$^{+0.91}_{-7.16}$ & 2.12$\pm$0.02 & 0.01$\pm$0.00 & star & 1&\cite{schulze11} \\ 
NGC2787 & 7.4$\pm$1.2 & 7.61$\pm$0.26 & 2.28$\pm$0.02 & 0.00$\pm$0.00 & gas & 1&\cite{sarzi01} \\ 
NGC2892 & 86.2$\pm$8.6 & 8.43$\pm$0.33 & 2.47$\pm$0.03 & 0.02$\pm$0.00 & gas & 1&\cite{beifiori12} \\ 
NGC2903 & 10.4$\pm$1.0 & 7.06$^{+0.85}_{-7.06}$ & 1.97$\pm$0.06 & 0.00$\pm$0.00 & gas & 1&\cite{beifiori12} \\ 
NGC2960 & 67.1$\pm$7.1 & 7.03$\pm$0.15 & 2.18$\pm$0.02 & 0.02$\pm$0.00 & maser & 1&\cite{kuo11,vdb15} \\ 
NGC2964 & 19.7$\pm$2.0 & 6.73$^{+1.84}_{-6.73}$ & 1.97$\pm$0.09 & 0.00$\pm$0.00 & gas & 1&\cite{beifiori12} \\ 
NGC2974 & 21.5$\pm$2.4 & 8.23$\pm$0.27 & 2.36$\pm$0.02 & 0.01$\pm$0.00 & star & 1&\cite{cappellari08} \\ 
NGC3021 & 22.4$\pm$2.2 & 7.26$^{+0.91}_{-7.26}$ & 1.70$\pm$0.20 & 0.01$\pm$0.00 & gas & 1&\cite{beifiori12} \\ 
NGC3031 & 3.6$\pm$0.1 & 7.81$\pm$0.39 & 2.15$\pm$0.02 & -0.00$\pm$0.00 & gas & 1&\cite{devereux03} \\ 
NGC3078 & 32.8$\pm$3.3 & 7.91$\pm$1.25 & 2.32$\pm$0.03 & 0.01$\pm$0.00 & gas & 1&\cite{beifiori12} \\ 
NGC3079 & 15.9$\pm$1.2 & 6.40$\pm$0.15 & 2.16$\pm$0.02 & 0.00$\pm$0.00 & maser & 1&\cite{trotter98,yamauchi04,kondratko05} \\ 
NGC3081 & 33.5$\pm$3.4 & 7.20$\pm$0.91 & 2.09$\pm$0.03 & 0.01$\pm$0.00 & gas & 1&\cite{beifiori12} \\ 
NGC3091 & 51.2$\pm$8.3 & 9.56$\pm$0.22 & 2.49$\pm$0.02 & 0.01$\pm$0.00 & star & 1&\cite{rusli13} \\ 
NGC3115 & 9.5$\pm$0.4 & 8.95$\pm$0.28 & 2.36$\pm$0.02 & 0.00$\pm$0.00 & star & 1&\cite{emsellem99} \\ 
NGC3227 & 23.8$\pm$2.6 & 7.32$\pm$0.70 & 2.12$\pm$0.04 & 0.00$\pm$0.00 & star & 1&\cite{davies06} \\ 
NGC3245 & 21.4$\pm$2.0 & 8.38$\pm$0.34 & 2.25$\pm$0.02 & 0.00$\pm$0.00 & gas & 1&\cite{barth01} \\ 
NGC3310 & 17.4$\pm$1.7 & 6.70$^{+2.77}_{-6.70}$ & 1.92$\pm$0.02 & 0.00$\pm$0.00 & gas & 1&\cite{pastorini07} \\ 
NGC3351 & 9.3$\pm$0.9 & 6.52$^{+0.78}_{-6.52}$ & 1.97$\pm$0.07 & 0.00$\pm$0.00 & gas & 1&\cite{beifiori12} \\ 
NGC3368 & 10.4$\pm$1.0 & 6.88$\pm$0.23 & 2.08$\pm$0.09 & 0.00$\pm$0.00 & star & 1&\cite{nowak10} \\ 
NGC3377 & 11.0$\pm$0.5 & 8.25$\pm$0.76 & 2.11$\pm$0.02 & 0.00$\pm$0.00 & star & 1&\cite{schulze11} \\ 
NGC3379 & 10.7$\pm$0.5 & 8.62$\pm$0.34 & 2.27$\pm$0.02 & 0.00$\pm$0.00 & star & 1&\cite{vdb10} \\ 
NGC3384 & 11.5$\pm$0.7 & 7.03$\pm$0.64 & 2.14$\pm$0.02 & 0.00$\pm$0.00 & star & 1&\cite{schulze11} \\ 
NGC3393 & 49.2$\pm$8.2 & 7.20$\pm$0.99 & 2.17$\pm$0.03 & 0.01$\pm$0.00 & maser & 1&\cite{kondratko08} \\ 
NGC3414 & 25.2$\pm$2.7 & 8.40$\pm$0.21 & 2.28$\pm$0.02 & 0.00$\pm$0.00 & star & 1&\cite{cappellari08} \\ 
NGC3423 & 14.6$\pm$1.5 & 5.18$^{+0.67}_{-5.18}$ & 1.62$\pm$0.07 & 0.00$\pm$0.00 & star & 3&\cite{neumayer12} \\ 
NGC3489 & 12.1$\pm$0.8 & 6.78$\pm$0.15 & 2.01$\pm$0.02 & 0.00$\pm$0.00 & star & 1&\cite{nowak10} \\ 
NGC3585 & 20.5$\pm$1.7 & 8.52$\pm$0.38 & 2.33$\pm$0.02 & 0.00$\pm$0.00 & star & 1&\cite{gultekin09} \\ 
NGC3607 & 22.6$\pm$1.8 & 8.14$\pm$0.47 & 2.32$\pm$0.02 & 0.00$\pm$0.00 & star & 1&\cite{gultekin09} \\ 
NGC3608 & 22.8$\pm$1.5 & 8.67$\pm$0.29 & 2.23$\pm$0.02 & 0.00$\pm$0.00 & star & 1&\cite{schulze11} \\ 
NGC3621 & 6.6$\pm$0.7 & 6.00$^{+1.43}_{-6.00}$ & 1.79$\pm$0.03 & 0.00$\pm$0.00 & star & 1&\cite{barth09} \\ 
NGC3627 & 10.1$\pm$1.1 & 6.93$\pm$0.14 & 2.09$\pm$0.01 & 0.00$\pm$0.00 & star & 1&\cite{saglia16} \\ 
NGC3642 & 21.6$\pm$2.2 & 7.42$^{+0.12}_{-7.42}$ & 1.97$\pm$0.11 & 0.01$\pm$0.00 & gas & 1&\cite{beifiori12} \\ 
NGC3665 & 34.7$\pm$6.7 & 8.76$\pm$0.27 & 2.34$\pm$0.02 & 0.01$\pm$0.00 & CO & 1&\cite{onishi16} \\ 
NGC3675 & 12.4$\pm$1.2 & 7.26$^{+0.88}_{-7.26}$ & 2.02$\pm$0.02 & 0.00$\pm$0.00 & gas & 1&\cite{beifiori12} \\ 
NGC3706 & 46.0$\pm$4.6 & 9.77$\pm$0.18 & 2.51$\pm$0.01 & 0.01$\pm$0.00 & star & 1&\cite{gultekin14} \\ 
NGC3783 & 41.7$\pm$4.2 & 7.36$\pm$0.57 & 1.98$\pm$0.05 & 0.01$\pm$0.00 & reverb & 1&\cite{onken02} \\ 
NGC3801 & 46.3$\pm$4.6 & 8.28$\pm$0.93 & 2.32$\pm$0.04 & 0.01$\pm$0.00 & gas & 1&\cite{beifiori12} \\ 
NGC3842 & 92.2$\pm$10.6 & 9.96$\pm$0.42 & 2.44$\pm$0.02 & 0.02$\pm$0.00 & star & 1&\cite{mcconnell11a} \\ 
NGC3862 & 84.6$\pm$8.5 & 8.41$\pm$1.11 & 2.32$\pm$0.03 & 0.02$\pm$0.00 & gas & 1&\cite{beifiori12} \\ 
NGC3923 & 20.9$\pm$2.7 & 9.45$\pm$0.35 & 2.35$\pm$0.02 & 0.01$\pm$0.00 & star & 1&\cite{saglia16} \\ 
NGC3945 & 19.5$\pm$2.0 & 6.94$^{+1.39}_{-6.94}$ & 2.25$\pm$0.02 & 0.00$\pm$0.00 & star & 1&\cite{gultekin09} \\ 
NGC3953 & 15.4$\pm$1.5 & 7.33$\pm$0.87 & 2.11$\pm$0.01 & 0.00$\pm$0.00 & gas & 1&\cite{beifiori12} \\ 
NGC3982 & 15.9$\pm$1.6 & 6.95$^{+0.77}_{-6.95}$ & 1.89$\pm$0.01 & 0.00$\pm$0.00 & gas & 1&\cite{beifiori12} \\ 
NGC3992 & 15.3$\pm$1.5 & 7.51$\pm$0.85 & 2.09$\pm$0.06 & 0.00$\pm$0.00 & gas & 1&\cite{beifiori12} \\ 
NGC3998 & 14.3$\pm$1.3 & 8.93$\pm$0.05 & 2.35$\pm$0.02 & 0.00$\pm$0.00 & star & 3&\cite{walsh12} \\ 
NGC4026 & 13.4$\pm$1.7 & 8.26$\pm$0.18 & 2.19$\pm$0.02 & 0.00$\pm$0.00 & star & 2&\cite{gultekin09} \\ 
NGC4036 & 19.0$\pm$1.9 & 7.89$\pm$1.09 & 2.26$\pm$0.02 & 0.00$\pm$0.00 & gas & 1&\cite{beifiori12} \\ 
NGC4051 & 10.0$\pm$1.0 & 6.10$\pm$0.75 & 1.95$\pm$0.01 & 0.00$\pm$0.00 & reverb & 1&\cite{denney09a} \\ 
NGC4088 & 11.9$\pm$1.2 & 6.79$^{+0.86}_{-6.79}$ & 1.93$\pm$0.02 & 0.00$\pm$0.00 & gas & 1&\cite{beifiori12} \\ 
NGC4143 & 14.8$\pm$1.5 & 7.92$\pm$1.07 & 2.25$\pm$0.02 & 0.00$\pm$0.00 & gas & 1&\cite{beifiori12} \\ 
NGC4150 & 12.8$\pm$1.3 & 5.94$^{+1.31}_{-5.94}$ & 1.91$\pm$0.02 & 0.00$\pm$0.00 & gas & 1&\cite{beifiori12} \\ 
NGC4151 & 20.0$\pm$2.8 & 7.81$\pm$0.23 & 1.98$\pm$0.01 & 0.00$\pm$0.00 & star & 1&\cite{onken14} \\ 
NGC4203 & 14.1$\pm$1.4 & 7.82$\pm$0.78 & 2.11$\pm$0.02 & 0.00$\pm$0.00 & gas & 1&\cite{beifiori12} \\ 
NGC4212 & 3.2$\pm$0.3 & 5.99$^{+1.27}_{-5.99}$ & 1.83$\pm$0.02 & -0.00$\pm$0.00 & gas & 1&\cite{beifiori12} \\ 
NGC4245 & 14.6$\pm$1.5 & 7.19$^{+1.43}_{-7.19}$ & 1.92$\pm$0.02 & 0.00$\pm$0.00 & gas & 1&\cite{beifiori12} \\ 
NGC4253 & 55.4$\pm$5.5 & 6.80$\pm$0.50 & 1.94$\pm$0.16 & 0.01$\pm$0.00 & reverb & 1&\cite{bentz10} \\ 
NGC4258 & 7.3$\pm$0.5 & 7.58$\pm$0.09 & 2.06$\pm$0.04 & 0.00$\pm$0.00 & maser & 1&\cite{herrnstein05} \\ 
NGC4261 & 32.4$\pm$2.8 & 8.72$\pm$0.29 & 2.42$\pm$0.02 & 0.01$\pm$0.00 & gas & 1&\cite{ferrarese96} \\ 
NGC4278 & 15.0$\pm$1.5 & 7.96$\pm$0.81 & 2.33$\pm$0.02 & 0.00$\pm$0.00 & gas & 1&\cite{beifiori12} \\ 
NGC4291 & 26.6$\pm$3.9 & 8.99$\pm$0.47 & 2.38$\pm$0.02 & 0.01$\pm$0.00 & star & 1&\cite{schulze11} \\ 
NGC4303 & 17.9$\pm$1.8 & 6.51$\pm$2.21 & 1.92$\pm$0.02 & 0.01$\pm$0.00 & gas & 1&\cite{pastorini07} \\ 
NGC4314 & 15.5$\pm$1.5 & 6.91$^{+0.89}_{-6.91}$ & 2.03$\pm$0.01 & 0.00$\pm$0.00 & gas & 1&\cite{beifiori12} \\ 
NGC4321 & 14.2$\pm$1.4 & 6.67$^{+0.50}_{-6.67}$ & 1.92$\pm$0.02 & 0.01$\pm$0.00 & gas & 1&\cite{beifiori12} \\ 
NGC4335 & 59.1$\pm$5.9 & 8.39$\pm$0.94 & 2.41$\pm$0.01 & 0.02$\pm$0.00 & gas & 1&\cite{beifiori12} \\ 
NGC4342 & 22.9$\pm$1.4 & 8.66$\pm$0.56 & 2.38$\pm$0.02 & 0.00$\pm$0.00 & star & 1&\cite{cretton99} \\ 
NGC4350 & 17.0$\pm$1.7 & 8.58$\pm$1.22 & 2.24$\pm$0.02 & 0.00$\pm$0.00 & star & 1&\cite{pignatelli01} \\ 
NGC4371 & 16.9$\pm$1.5 & 6.84$\pm$0.22 & 2.16$\pm$0.02 & 0.00$\pm$0.00 & star & 1&\cite{saglia16} \\ 
NGC4374 & 18.5$\pm$0.6 & 8.97$\pm$0.14 & 2.41$\pm$0.02 & 0.00$\pm$0.00 & gas & 1&\cite{walsh10} \\ 
NGC4382 & 17.9$\pm$1.8 & 7.11$^{+3.71}_{-7.11}$ & 2.25$\pm$0.02 & 0.00$\pm$0.00 & star & 1&\cite{gultekin11} \\ 
NGC4388 & 16.5$\pm$1.6 & 6.86$\pm$0.13 & 2.03$\pm$0.03 & 0.01$\pm$0.00 & maser & 1&\cite{kuo11} \\ 
NGC4429 & 18.2$\pm$1.8 & 7.85$\pm$1.05 & 2.25$\pm$0.02 & 0.00$\pm$0.00 & gas & 1&\cite{beifiori12} \\ 
NGC4434 & 20.8$\pm$1.5 & 7.85$\pm$0.54 & 1.99$\pm$1.04 & 0.00$\pm$0.00 & star & 1&\cite{graham18} \\ 
NGC4435 & 17.0$\pm$1.7 & 0.00$^{+6.30}_{-0.00}$ & 2.18$\pm$0.02 & 0.00$\pm$0.00 & star & 3&\cite{coccato06} \\ 
NGC4450 & 28.3$\pm$2.8 & 8.06$^{+0.84}_{-8.06}$ & 2.03$\pm$0.06 & 0.01$\pm$0.00 & gas & 1&\cite{beifiori12} \\ 
NGC4459 & 16.0$\pm$0.5 & 7.84$\pm$0.26 & 2.20$\pm$0.02 & 0.00$\pm$0.00 & gas & 1&\cite{sarzi01} \\ 
NGC4472 & 17.1$\pm$0.6 & 9.40$\pm$0.11 & 2.40$\pm$0.02 & 0.00$\pm$0.00 & star & 1&\cite{rusli13} \\ 
NGC4473 & 15.2$\pm$0.5 & 7.95$\pm$0.72 & 2.27$\pm$0.02 & 0.01$\pm$0.00 & star & 1&\cite{schulze11} \\ 
NGC4477 & 20.8$\pm$2.1 & 7.55$\pm$0.89 & 2.17$\pm$0.02 & 0.00$\pm$0.00 & gas & 1&\cite{beifiori12} \\ 
NGC4486 & 16.7$\pm$0.6 & 9.58$\pm$0.29 & 2.42$\pm$0.02 & 0.00$\pm$0.00 & gas & 1&\cite{walsh13} \\ 
NGC4486A & 16.0$\pm$0.5 & 7.10$\pm$0.44 & 2.09$\pm$0.02 & 0.00$\pm$0.00 & star & 1&\cite{nowak07} \\ 
NGC4486B & 16.5$\pm$0.6 & 8.60$\pm$0.07 & 2.23$\pm$0.02 & 0.01$\pm$0.00 & star & 1&\cite{saglia16} \\ 
NGC4501 & 16.5$\pm$1.1 & 7.30$\pm$0.24 & 2.20$\pm$0.01 & 0.01$\pm$0.00 & star & 1&\cite{saglia16} \\ 
NGC4507 & 47.0$\pm$4.7 & 7.18$\pm$1.05 & 2.16$\pm$0.02 & 0.01$\pm$0.00 & gas & 1&\cite{beifiori12} \\ 
NGC4526 & 16.4$\pm$1.8 & 8.65$\pm$0.37 & 2.32$\pm$0.02 & 0.00$\pm$0.00 & gas & 1&\cite{davis13a} \\ 
NGC4548 & 17.9$\pm$1.8 & 7.25$\pm$0.88 & 2.16$\pm$0.04 & 0.00$\pm$0.00 & gas & 1&\cite{beifiori12} \\ 
NGC4552 & 15.3$\pm$1.0 & 8.70$\pm$0.15 & 2.35$\pm$0.02 & 0.00$\pm$0.00 & star & 1&\cite{cappellari08} \\ 
NGC4564 & 15.9$\pm$0.5 & 7.95$\pm$0.37 & 2.19$\pm$0.02 & 0.00$\pm$0.00 & star & 1&\cite{schulze11} \\ 
NGC4578 & 38.7$\pm$2.7 & 7.28$\pm$1.23 & 2.03$\pm$0.95 & 0.01$\pm$0.00 & star & 1&\cite{graham18} \\ 
NGC4579 & 23.0$\pm$2.3 & 7.96$^{+1.08}_{-7.96}$ & 2.04$\pm$0.06 & 0.01$\pm$0.00 & gas & 1&\cite{beifiori12} \\ 
NGC4593 & 38.5$\pm$3.9 & 6.86$\pm$0.62 & 2.13$\pm$0.02 & 0.01$\pm$0.00 & reverb & 1&\cite{barth13} \\ 
NGC4594 & 9.9$\pm$0.8 & 8.82$\pm$0.14 & 2.38$\pm$0.02 & 0.00$\pm$0.00 & star & 1&\cite{jardel11} \\ 
NGC4596 & 16.5$\pm$6.2 & 7.89$\pm$0.78 & 2.10$\pm$0.02 & 0.01$\pm$0.00 & gas & 1&\cite{sarzi01} \\ 
NGC4621 & 18.3$\pm$3.0 & 8.60$\pm$0.26 & 2.30$\pm$0.02 & 0.00$\pm$0.00 & star & 1&\cite{cappellari08} \\ 
NGC4649 & 16.5$\pm$0.6 & 9.67$\pm$0.30 & 2.43$\pm$0.02 & 0.00$\pm$0.00 & star & 1&\cite{shen10} \\ 
NGC4697 & 12.5$\pm$0.4 & 8.31$\pm$0.34 & 2.23$\pm$0.02 & 0.00$\pm$0.00 & star & 1&\cite{schulze11} \\ 
NGC4698 & 16.4$\pm$1.6 & 7.76$\pm$0.48 & 2.09$\pm$0.03 & 0.00$\pm$0.00 & gas & 1&\cite{beifiori12} \\ 
NGC4742 & 15.7$\pm$1.6 & 7.10$\pm$0.45 & 1.95$\pm$0.02 & 0.00$\pm$0.00 & star & 1&\cite{tremaine02} \\ 
NGC4748 & 62.7$\pm$6.3 & 6.36$\pm$0.78 & 2.02$\pm$0.05 & 0.01$\pm$0.00 & reverb & 1&\cite{bentz10} \\ 
NGC4751 & 26.9$\pm$2.9 & 9.15$\pm$0.17 & 2.55$\pm$0.02 & 0.01$\pm$0.00 & star & 1&\cite{rusli13} \\ 
NGC4762 & 19.2$\pm$1.4 & 7.36$\pm$0.54 & 2.13$\pm$0.95 & 0.00$\pm$0.00 & star & 1&\cite{graham18} \\ 
NGC4800 & 12.5$\pm$1.3 & 7.02$^{+1.60}_{-7.02}$ & 2.01$\pm$0.01 & 0.00$\pm$0.00 & gas & 1&\cite{beifiori12} \\ 
NGC4889 & 102.0$\pm$5.2 & 10.32$\pm$1.31 & 2.56$\pm$0.02 & 0.02$\pm$0.00 & star & 1&\cite{mcconnell11a} \\ 
NGC4945 & 3.7$\pm$0.4 & 6.14$\pm$0.55 & 2.13$\pm$0.02 & 0.00$\pm$0.00 & maser & 1&\cite{greenhill97} \\ 
NGC5005 & 14.6$\pm$1.5 & 8.27$\pm$0.70 & 2.30$\pm$0.02 & 0.00$\pm$0.00 & gas & 1&\cite{beifiori12} \\ 
NGC5055 & 8.7$\pm$0.9 & 8.92$\pm$0.30 & 2.00$\pm$0.02 & 0.00$\pm$0.00 & gas & 1&\cite{blais-ouellette04} \\ 
NGC5077 & 38.7$\pm$8.4 & 8.93$\pm$0.80 & 2.35$\pm$0.02 & 0.01$\pm$0.00 & gas & 1&\cite{de-francesco08} \\ 
NGC5127 & 62.5$\pm$6.3 & 8.27$\pm$1.24 & 2.29$\pm$0.11 & 0.02$\pm$0.00 & gas & 1&\cite{beifiori12} \\ 
NGC5128 & 3.6$\pm$0.2 & 7.76$\pm$0.25 & 2.18$\pm$0.02 & 0.00$\pm$0.00 & star & 1&\cite{cappellari09} \\ 
NGC5248 & 17.9$\pm$1.8 & 6.30$\pm$1.14 & 2.10$\pm$0.04 & 0.00$\pm$0.00 & gas & 1&\cite{beifiori12} \\ 
NGC5252 & 103.7$\pm$10.4 & 9.07$\pm$1.02 & 2.28$\pm$0.02 & 0.02$\pm$0.00 & gas & 1&\cite{capetti05} \\ 
NGC5273 & 15.5$\pm$1.5 & 6.61$\pm$0.81 & 1.82$\pm$0.02 & 0.00$\pm$0.00 & reverb & 1&\cite{bentz14,vdb15} \\ 
NGC5283 & 34.5$\pm$3.5 & 7.41$\pm$0.99 & 2.13$\pm$0.04 & 0.10$\pm$0.00 & gas & 1&\cite{beifiori12} \\ 
NGC5328 & 64.1$\pm$7.0 & 9.67$\pm$0.16 & 2.52$\pm$0.02 & 0.02$\pm$0.00 & star & 3&\cite{rusli13} \\ 
NGC5347 & 32.3$\pm$3.2 & 7.21$^{+1.27}_{-7.21}$ & 1.80$\pm$0.08 & 0.01$\pm$0.00 & gas & 1&\cite{beifiori12} \\ 
NGC5419 & 56.2$\pm$6.1 & 9.86$\pm$0.43 & 2.57$\pm$0.01 & 0.01$\pm$0.00 & star & 1&\cite{saglia16} \\ 
NGC5427 & 36.3$\pm$3.6 & 7.58$^{+0.90}_{-7.58}$ & 1.80$\pm$0.08 & 0.01$\pm$0.00 & gas & 1&\cite{beifiori12} \\ 
NGC5457 & 7.0$\pm$0.7 & 6.41$^{+0.23}_{-6.41}$ & 1.35$\pm$0.17 & 0.00$\pm$0.00 & star & 1&\cite{kormendy10} \\ 
NGC5490 & 65.2$\pm$6.5 & 8.73$\pm$1.05 & 2.41$\pm$0.04 & 0.02$\pm$0.00 & gas & 1&\cite{beifiori12} \\ 
NGC5495 & 126.3$\pm$11.6 & 7.00$\pm$0.15 & 2.22$\pm$0.05 & 0.02$\pm$0.00 & maser & 1&\cite{greene16} \\ 
NGC5516 & 58.4$\pm$6.3 & 9.52$\pm$0.17 & 2.48$\pm$0.04 & 0.01$\pm$0.00 & star & 1&\cite{rusli13} \\ 
NGC5548 & 73.6$\pm$7.4 & 7.70$\pm$0.39 & 2.29$\pm$0.03 & 0.02$\pm$0.00 & reverb & 1&\cite{kovacevic14} \\ 
NGC5576 & 25.7$\pm$1.7 & 8.44$\pm$0.38 & 2.19$\pm$0.02 & 0.01$\pm$0.00 & star & 1&\cite{gultekin09} \\ 
NGC5643 & 17.4$\pm$1.7 & 7.05$^{+1.77}_{-7.05}$ & 1.95$\pm$0.01 & 0.00$\pm$0.00 & gas & 1&\cite{beifiori12} \\ 
NGC5695 & 54.6$\pm$5.5 & 8.00$\pm$0.96 & 2.16$\pm$0.01 & 0.01$\pm$0.00 & gas & 1&\cite{beifiori12} \\ 
NGC5728 & 37.6$\pm$3.8 & 8.05$\pm$0.87 & 2.28$\pm$0.09 & 0.01$\pm$0.00 & gas & 1&\cite{beifiori12} \\ 
NGC5765B & 113.0$\pm$11.3 & 7.66$\pm$0.09 & 2.20$\pm$0.05 & 0.03$\pm$0.00 & maser & 1&\cite{gao15} \\ 
NGC5813 & 32.2$\pm$2.7 & 8.85$\pm$0.17 & 2.32$\pm$0.02 & 0.01$\pm$0.00 & star & 1&\cite{cappellari08} \\ 
NGC5845 & 25.9$\pm$4.1 & 8.69$\pm$0.47 & 2.36$\pm$0.02 & 0.01$\pm$0.00 & star & 1&\cite{schulze11} \\ 
NGC5846 & 24.9$\pm$2.3 & 9.04$\pm$0.17 & 2.35$\pm$0.02 & 0.01$\pm$0.00 & star & 1&\cite{cappellari08} \\ 
NGC5879 & 10.6$\pm$1.1 & 6.62$^{+0.83}_{-6.62}$ & 1.73$\pm$0.06 & 0.00$\pm$0.00 & gas & 1&\cite{beifiori12} \\ 
NGC5921 & 20.5$\pm$2.0 & 7.07$^{+1.27}_{-7.07}$ & 1.93$\pm$0.05 & 0.00$\pm$0.00 & gas & 1&\cite{beifiori12} \\ 
NGC6086 & 138.0$\pm$11.5 & 9.57$\pm$0.50 & 2.50$\pm$0.02 & 0.03$\pm$0.00 & star & 1&\cite{mcconnell11b} \\ 
NGC6240(S) & 105.0$\pm$10.5 & 9.17$\pm$0.64 & 2.44$\pm$0.07 & 0.02$\pm$0.00 & star & 1&\cite{medling11} \\ 
NGC6251 & 108.4$\pm$9.0 & 8.79$\pm$0.47 & 2.46$\pm$0.02 & 0.02$\pm$0.00 & gas & 1&\cite{ferrarese99} \\ 
NGC6300 & 14.2$\pm$1.4 & 7.14$^{+0.60}_{-7.14}$ & 1.94$\pm$0.02 & 0.00$\pm$0.00 & gas & 1&\cite{beifiori12} \\ 
NGC6323 & 113.4$\pm$12.3 & 7.00$\pm$0.14 & 2.20$\pm$0.07 & 0.03$\pm$0.00 & maser & 1&\cite{kuo11} \\ 
NGC6500 & 36.5$\pm$3.6 & 8.34$\pm$0.78 & 2.33$\pm$0.01 & 0.01$\pm$0.00 & gas & 1&\cite{beifiori12} \\ 
NGC6503 & 5.3$\pm$0.5 & 6.30$^{+0.34}_{-6.30}$ & 1.74$\pm$0.02 & 0.00$\pm$0.00 & star & 1&\cite{kormendy10} \\ 
NGC6814 & 22.3$\pm$2.2 & 7.02$\pm$0.51 & 1.98$\pm$0.01 & 0.01$\pm$0.00 & reverb & 1&\cite{bentz10} \\ 
NGC6861 & 27.3$\pm$4.5 & 9.30$\pm$0.25 & 2.59$\pm$0.02 & 0.01$\pm$0.00 & star & 1&\cite{rusli13} \\ 
NGC6951 & 16.0$\pm$1.6 & 6.93$^{+0.56}_{-6.93}$ & 1.98$\pm$0.04 & 0.00$\pm$0.00 & gas & 1&\cite{beifiori12} \\ 
NGC7052 & 70.4$\pm$8.4 & 8.60$\pm$0.69 & 2.42$\pm$0.02 & 0.02$\pm$0.00 & gas & 1&\cite{van-der-marel98} \\ 
NGC7331 & 12.2$\pm$1.2 & 8.02$\pm$0.55 & 2.06$\pm$0.02 & 0.00$\pm$0.00 & gas & 1&\cite{beifiori12} \\ 
NGC7332 & 21.7$\pm$2.2 & 7.08$\pm$0.53 & 2.10$\pm$0.02 & 0.00$\pm$0.00 & star & 1&\cite{haring04} \\ 
NGC7418 & 18.4$\pm$1.8 & 5.18$^{+1.78}_{-5.18}$ & 1.67$\pm$0.06 & 0.00$\pm$0.00 & star & 3&\cite{neumayer12} \\ 
NGC7424 & 10.9$\pm$1.1 & 5.18$^{+0.43}_{-5.18}$ & 1.34$\pm$0.05 & 0.00$\pm$0.00 & star & 3&\cite{neumayer12} \\ 
NGC7457 & 12.5$\pm$1.2 & 6.95$\pm$0.91 & 1.87$\pm$0.02 & 0.00$\pm$0.00 & star & 1&\cite{schulze11} \\ 
NGC7582 & 22.3$\pm$9.8 & 7.74$\pm$0.61 & 2.19$\pm$0.05 & 0.01$\pm$0.00 & gas & 1&\cite{wold06} \\ 
NGC7619 & 51.5$\pm$7.4 & 9.40$\pm$0.32 & 2.51$\pm$0.02 & 0.01$\pm$0.00 & star & 1&\cite{rusli13} \\ 
NGC7626 & 38.1$\pm$3.8 & 8.58$\pm$0.98 & 2.37$\pm$0.02 & 0.01$\pm$0.00 & gas & 1&\cite{beifiori12} \\ 
NGC7682 & 59.3$\pm$5.9 & 7.56$\pm$0.98 & 2.05$\pm$0.06 & 0.02$\pm$0.00 & gas & 1&\cite{beifiori12} \\ 
NGC7768 & 116.0$\pm$27.5 & 9.13$\pm$0.54 & 2.42$\pm$0.02 & 0.03$\pm$0.00 & star & 1&\cite{mcconnell12} \\ 
NSA10779 & 208.2$\pm$14.6 & 5.44$\pm$1.20 & 1.53$\pm$0.78 & 0.05$\pm$0.00 & reverb & 1&\cite{baldassare20} \\ 
NSA125318 & 147.4$\pm$10.3 & 5.00$\pm$1.20 & 1.61$\pm$0.78 & 0.03$\pm$0.00 & reverb & 1&\cite{baldassare20} \\ 
NSA15235 & 135.3$\pm$9.5 & 5.29$\pm$1.20 & 1.62$\pm$1.15 & 0.03$\pm$0.00 & reverb & 1&\cite{baldassare20} \\ 
NSA166155 & 109.8$\pm$7.7 & 4.70$\pm$1.20 & 1.45$\pm$1.05 & 0.02$\pm$0.00 & reverb & 1&\cite{baldassare15} \\ 
NSA18913 & 176.9$\pm$12.4 & 5.18$\pm$1.20 & 1.52$\pm$1.15 & 0.04$\pm$0.00 & reverb & 1&\cite{baldassare20} \\ 
NSA47066 & 184.3$\pm$12.9 & 5.42$\pm$1.20 & 1.51$\pm$0.70 & 0.04$\pm$0.00 & reverb & 1&\cite{baldassare20} \\ 
NSA52675 & 70.2$\pm$4.9 & 6.10$\pm$1.20 & 1.72$\pm$0.70 & 0.02$\pm$0.00 & reverb & 1&\cite{baldassare20} \\ 
NSA62996 & 199.9$\pm$14.0 & 5.80$\pm$1.20 & 1.82$\pm$0.48 & 0.05$\pm$0.00 & reverb & 1&\cite{baldassare20} \\ 
NSA79874 & 172.2$\pm$12.1 & 5.46$\pm$1.20 & 1.45$\pm$0.85 & 0.04$\pm$0.00 & reverb & 1&\cite{baldassare16} \\ 
NSA99052 & 141.2$\pm$9.9 & 5.21$\pm$1.20 & 1.72$\pm$0.70 & 0.03$\pm$0.00 & reverb & 1&\cite{baldassare20} \\ 
SBS1116+583A & 119.4$\pm$11.9 & 6.54$\pm$0.59 & 1.96$\pm$0.02 & 0.03$\pm$0.00 & reverb & 1&\cite{bentz10} \\ 
UGC12064 & 34.7$\pm$6.7 & 8.84$\pm$1.56 & 2.41$\pm$0.03 & 0.02$\pm$0.00 & gas & 1&\cite{beifiori12} \\ 
UGC1214 & 59.9$\pm$6.0 & 7.74$^{+0.54}_{-7.74}$ & 2.02$\pm$0.06 & 0.02$\pm$0.00 & gas & 1&\cite{beifiori12} \\ 
UGC1395 & 60.8$\pm$6.1 & 6.83$^{+0.85}_{-6.83}$ & 1.81$\pm$0.04 & 0.02$\pm$0.00 & gas & 1&\cite{beifiori12} \\ 
UGC1841 & 74.9$\pm$7.5 & 8.47$\pm$0.58 & 2.47$\pm$0.04 & 0.02$\pm$0.00 & gas & 1&\cite{beifiori12} \\ 
UGC3789 & 49.9$\pm$5.4 & 6.99$\pm$0.26 & 2.03$\pm$0.05 & 0.01$\pm$0.00 & maser & 1&\cite{kuo11} \\ 
UGC6093 & 150.0$\pm$15.0 & 7.41$\pm$0.06 & 2.19$\pm$0.05 & 0.04$\pm$0.00 & maser & 1&\cite{greene16} \\ 
UGC7115 & 88.2$\pm$8.8 & 9.00$\pm$1.34 & 2.25$\pm$0.08 & 0.02$\pm$0.00 & gas & 1&\cite{beifiori12} \\ 
UGC9799 & 151.1$\pm$15.1 & 8.89$^{+2.41}_{-8.89}$ & 2.37$\pm$0.02 & 0.03$\pm$0.00 & gas & 1&\cite{bonta09} \\ 
\enddata
\tablecomments{(1) Name. (2) Distance. (3) BH mass. (4) Effective stellar velocity dispersion. (5) Method used for BH mass measurement. BH masses measured using stellar dynamics, gas dynamics, reverberation mapping and water masers are denoted by keywords star, gas, reverb, maser, respectively. (6) $1\sigma_s$, $2\sigma_s$ or $3\sigma_s$ confidence level on $M_\bullet$ uncertainty. (7) Literature reference.}
\label{table:included}
\end{deluxetable*}
\end{longrotatetable}

\startlongtable
\begin{longrotatetable}
\begin{deluxetable*}{lrrrhrcl}
\tablecaption{Sample of the Black Hole Mass Measurements Excluded from the Fits\label{chartable}}
\tablewidth{700pt}
\tabletypesize{\scriptsize}
\tablehead{
\colhead{Name} & \colhead{Distance} & 
\colhead{$M_\bullet$} & \colhead{$\sigma_e$} & \colhead{} &
\colhead{Method} & \colhead{$M_\bullet$ Uncertainty} & \colhead{References}   \\ 
\colhead{} & \colhead{(Mpc)} & \colhead{log($M_\odot$)} & \colhead{log(km $s^{-1}$)} & 
\colhead{} & \colhead{} & \colhead{$1/2/3\sigma_s$} & \colhead{} 
} 
\decimalcolnumbers
\startdata
1RXSJ1858+4850 & 338.3$\pm$33.8 & 6.68$\pm$0.56 & -1234.00$\pm$-1234.00 & nan$\pm$nan & reverb&1 & \cite{pei14} \\ 
G1 & 0.7$\pm$0.1 & 5.24$\pm$0.37 & 1.39$\pm$0.09 & 0.26$\pm$0.00 & GC&1 & \cite{gebhardt05} \\ 
H0507+164 & 76.6$\pm$7.7 & 6.69$\pm$1.09 & -1234.00$\pm$-1234.00 & 0.02$\pm$0.00 & reverb&1 & \cite{stalin11} \\ 
M60UCD1 & 16.5$\pm$1.6 & 7.30$\pm$0.45 & 1.82$\pm$0.03 & nan$\pm$nan & star&1 & \cite{seth14} \\ 
Mrk0006 & 80.6$\pm$8.1 & 8.09$\pm$0.46 & -1234.00$\pm$-1234.00 & 0.02$\pm$0.00 & reverb&1 & \cite{doroshenko12} \\ 
Mrk0110 & 151.1$\pm$15.1 & 7.28$\pm$0.64 & 1.96$\pm$0.03 & 0.04$\pm$0.00 & reverb&1 & \cite{peterson04} \\ 
Mrk0142 & 192.5$\pm$19.2 & 6.27$\pm$0.63 & -1234.00$\pm$-1234.00 & 0.04$\pm$0.00 & reverb&1 & \cite{du14} \\ 
Mrk0279 & 130.4$\pm$13.0 & 7.40$\pm$0.69 & 2.29$\pm$0.03 & 0.03$\pm$0.00 & reverb&1 & \cite{peterson04} \\ 
Mrk0290 & 126.7$\pm$12.7 & 7.26$\pm$0.52 & -1234.00$\pm$-1234.00 & 0.03$\pm$0.00 & reverb&1 & \cite{denney10} \\ 
Mrk0335 & 110.5$\pm$11.0 & 7.21$\pm$0.47 & -1234.00$\pm$-1234.00 & 0.03$\pm$0.00 & reverb&1 & \cite{du14} \\ 
Mrk1029 & 124.0$\pm$12.4 & 6.28$\pm$0.39 & 2.12$\pm$0.05 & 0.03$\pm$0.00 & maser&3 & \cite{greene16} \\ 
Mrk1501 & 382.6$\pm$38.3 & 8.03$\pm$0.77 & -1234.00$\pm$-1234.00 & 0.09$\pm$0.00 & reverb&1 & \cite{grier12} \\ 
MW & 0.0$\pm$0.0 & 6.63$\pm$0.14 & 2.00$\pm$0.09 & nan$\pm$nan & star&1 & \cite{ghez08,gillessen09} \\ 
NGC0104 & 0.0$\pm$0.0 & 3.00$^{+0.53}_{-3.00}$ & 1.06$\pm$0.03 & nan$\pm$nan & GC&1 & \cite{mclaughlin06a} \\ 
NGC0224 & 0.8$\pm$0.0 & 8.15$\pm$0.48 & 2.20$\pm$0.02 & -0.00$\pm$0.00 & star&3 & \cite{bender05} \\ 
NGC0300 & 2.2$\pm$0.2 & 2.00$^{+3.00}_{-2.00}$ & 1.11$\pm$0.07 & 0.00$\pm$0.00 & star&1 & \cite{neumayer12} \\ 
NGC1068 & 15.9$\pm$9.4 & 6.92$\pm$0.74 & 2.18$\pm$0.02 & 0.00$\pm$0.00 & maser&1 & \cite{lodato03} \\ 
NGC1300 & 21.5$\pm$9.4 & 7.88$\pm$1.03 & 2.34$\pm$0.02 & 0.01$\pm$0.00 & gas&1 & \cite{atkinson05} \\ 
NGC1851 & 0.0$\pm$0.0 & 0.00$^{+3.30}_{-0.00}$ & 0.97$\pm$0.02 & nan$\pm$nan & GC&1 & \cite{lutzgendorf13} \\ 
NGC1904 & 0.0$\pm$0.0 & 3.48$^{+0.12}_{-3.48}$ & 0.90$\pm$0.03 & nan$\pm$nan & GC&1 & \cite{lutzgendorf13} \\ 
NGC2139 & 23.6$\pm$2.4 & 5.18$^{+0.43}_{-5.18}$ & 1.34$\pm$0.08 & 0.01$\pm$0.00 & star&1 & \cite{neumayer12} \\ 
NGC2911 & 43.5$\pm$4.3 & 9.09$\pm$0.88 & 2.32$\pm$0.03 & 0.01$\pm$0.00 & gas&1 & \cite{beifiori12} \\ 
NGC3227 & 23.8$\pm$2.6 & 6.75$\pm$0.21 & 1.96$\pm$0.03 & 0.00$\pm$0.00 & reverb&1 & \cite{denney10} \\ 
NGC3516 & 37.9$\pm$3.8 & 7.37$\pm$0.16 & 2.26$\pm$0.01 & 0.01$\pm$0.00 & reverb&1 & \cite{denney10} \\ 
NGC4041 & 19.5$\pm$2.0 & 6.00$^{+0.61}_{-6.00}$ & 1.98$\pm$0.02 & 0.00$\pm$0.00 & gas&1 & \cite{marconi03} \\ 
NGC4151 & 20.0$\pm$2.8 & 7.54$\pm$0.49 & 1.99$\pm$0.01 & 0.00$\pm$0.00 & reverb&1 & \cite{bentz06a} \\ 
NGC4244 & 4.4$\pm$0.4 & 0.00$^{+5.66}_{-0.00}$ & 1.45$\pm$0.15 & 0.00$\pm$0.00 & star&3 & \cite{de-lorenzi13} \\ 
NGC4395 & 4.4$\pm$0.4 & 5.43$\pm$0.75 & 1.47$\pm$0.07 & nan$\pm$nan & reverb&1 & \cite{peterson05} \\ 
NGC4395 & 4.4$\pm$0.4 & 5.54$\pm$1.62 & 1.47$\pm$0.07 & 0.00$\pm$0.00 & gas&1 & \cite{den-brok15} \\ 
NGC4636 & 13.7$\pm$1.4 & 8.58$\pm$0.66 & 2.26$\pm$0.02 & 0.00$\pm$0.00 & gas&1 & \cite{beifiori12} \\ 
NGC4699 & 18.9$\pm$2.1 & 8.25$\pm$0.16 & 2.26$\pm$0.01 & 0.00$\pm$0.00 & star&1 & \cite{saglia16} \\ 
NGC4736 & 4.5$\pm$0.8 & 6.78$\pm$0.37 & 2.05$\pm$0.02 & 0.00$\pm$0.00 & star&1 & \cite{kormendy11} \\ 
NGC4826 & 5.2$\pm$1.2 & 6.05$\pm$0.39 & 1.98$\pm$0.02 & 0.00$\pm$0.00 & star&1 & \cite{kormendy11} \\ 
NGC5018 & 40.5$\pm$4.9 & 8.02$\pm$0.23 & 2.32$\pm$0.01 & 0.01$\pm$0.00 & star&3 & \cite{saglia16} \\ 
NGC5139 & 0.0$\pm$0.0 & 3.94$^{+0.42}_{-3.94}$ & 1.28$\pm$0.03 & nan$\pm$nan & GC&1 & \cite{van-der-marel10} \\ 
NGC5194 & 7.9$\pm$0.8 & 5.96$^{+1.08}_{-5.96}$ & 1.84$\pm$0.06 & 0.00$\pm$0.00 & gas&1 & \cite{beifiori12} \\ 
NGC5272 & 0.0$\pm$0.0 & 0.00$^{+3.72}_{-0.00}$ & 0.73$\pm$0.08 & nan$\pm$nan & GC&1 & \cite{kamann14} \\ 
NGC5694 & 0.0$\pm$0.0 & 0.00$^{+3.90}_{-0.00}$ & 0.94$\pm$0.03 & -0.00$\pm$0.00 & GC&1 & \cite{lutzgendorf13} \\ 
NGC5793 & 53.3$\pm$5.3 & 7.30$\pm$7.30 & -1234.00$\pm$-1234.00 & 0.01$\pm$0.00 & maser&1 & \cite{hagiwara01} \\ 
NGC5824 & 0.0$\pm$0.0 & 0.00$^{+3.78}_{-0.00}$ & 1.05$\pm$0.02 & -0.00$\pm$0.00 & GC&1 & \cite{lutzgendorf13} \\ 
NGC6093 & 0.0$\pm$0.0 & 0.00$^{+2.90}_{-0.00}$ & 0.97$\pm$0.01 & nan$\pm$nan & GC&1 & \cite{lutzgendorf13} \\ 
NGC6205 & 0.0$\pm$0.0 & 0.00$^{+3.93}_{-0.00}$ & 0.85$\pm$0.06 & nan$\pm$nan & GC&1 & \cite{kamann14} \\ 
NGC6264 & 147.6$\pm$16.0 & 7.49$\pm$0.14 & 2.20$\pm$0.04 & 0.03$\pm$0.00 & maser&1 & \cite{kuo11} \\ 
NGC6266 & 0.0$\pm$0.0 & 3.30$^{+0.18}_{-3.30}$ & 1.19$\pm$0.01 & nan$\pm$nan & GC&3 & \cite{lutzgendorf13,mcnamara12} \\ 
NGC6341 & 0.0$\pm$0.0 & 0.00$^{+2.99}_{-0.00}$ & 0.77$\pm$0.07 & nan$\pm$nan & GC&1 & \cite{kamann14} \\ 
NGC6388 & 0.0$\pm$0.0 & 4.23$^{+0.55}_{-4.23}$ & 1.28$\pm$0.01 & 0.04$\pm$0.00 & GC&1 & \cite{lutzgendorf11} \\ 
NGC6397 & 0.0$\pm$0.0 & 2.78$^{+0.37}_{-2.78}$ & 0.69$\pm$0.09 & nan$\pm$nan & GC&1 & \cite{kamann16} \\ 
NGC7078 & 0.0$\pm$0.0 & 2.70$^{+2.33}_{-2.70}$ & 1.10$\pm$0.20 & nan$\pm$nan & GC&1 & \cite{vdb06,den-brok14} \\ 
NGC7469 & 47.7$\pm$8.1 & 6.94$\pm$0.16 & 2.12$\pm$0.02 & 0.02$\pm$0.00 & reverb&1 & \cite{peterson14} \\ 
NGC7793 & 3.3$\pm$0.3 & 3.70$^{+2.20}_{-3.70}$ & 1.39$\pm$0.07 & 0.00$\pm$0.00 & star&1 & \cite{neumayer12} \\ 
PG0026+129 & 608.2$\pm$60.8 & 8.46$\pm$0.66 & -1234.00$\pm$-1234.00 & 0.14$\pm$0.00 & reverb&1 & \cite{peterson04} \\ 
PG0052+251 & 661.5$\pm$66.1 & 8.44$\pm$0.61 & -1234.00$\pm$-1234.00 & 0.15$\pm$0.00 & reverb&1 & \cite{peterson04} \\ 
PG0804+761 & 428.3$\pm$42.8 & 8.72$\pm$0.16 & -1234.00$\pm$-1234.00 & 0.10$\pm$0.00 & reverb&1 & \cite{peterson04} \\ 
PG0953+414 & 1002.6$\pm$100.3 & 8.31$\pm$0.62 & -1234.00$\pm$-1234.00 & 0.23$\pm$0.00 & reverb&1 & \cite{peterson04} \\ 
PG1226+023 & 678.1$\pm$67.8 & 8.81$\pm$0.63 & -1234.00$\pm$-1234.00 & 0.16$\pm$0.00 & reverb&1 & \cite{peterson04} \\ 
PG1229+204 & 269.9$\pm$27.0 & 7.72$\pm$0.93 & 2.20$\pm$0.09 & 0.06$\pm$0.00 & reverb&1 & \cite{peterson04} \\ 
PG1307+085 & 663.8$\pm$66.4 & 8.49$\pm$0.72 & -1234.00$\pm$-1234.00 & 0.15$\pm$0.00 & reverb&1 & \cite{peterson04} \\ 
PG1411+442 & 383.7$\pm$38.4 & 8.50$\pm$0.78 & 2.32$\pm$0.06 & 0.09$\pm$0.00 & reverb&1 & \cite{peterson04} \\ 
PG1426+015 & 370.8$\pm$37.1 & 8.96$\pm$0.25 & 2.34$\pm$0.03 & 0.09$\pm$0.00 & reverb&1 & \cite{peterson04} \\ 
PG1613+658 & 552.5$\pm$55.2 & 8.27$\pm$0.33 & -1234.00$\pm$-1234.00 & 0.12$\pm$0.00 & reverb&3 & \cite{peterson04} \\ 
PG1617+175 & 481.6$\pm$48.2 & 8.63$\pm$0.66 & 2.30$\pm$0.08 & 0.11$\pm$0.00 & reverb&3 & \cite{peterson04} \\ 
PG1700+518 & 1250.6$\pm$125.1 & 8.76$\pm$0.63 & -1234.00$\pm$-1234.00 & 0.29$\pm$0.00 & reverb&1 & \cite{peterson04} \\ 
PG2130+099 & 269.7$\pm$27.0 & 7.41$\pm$0.52 & 2.21$\pm$0.05 & 0.06$\pm$0.00 & reverb&3 & \cite{grier12} \\ 
Zw229-015 & 119.4$\pm$11.9 & 6.88$\pm$0.63 & -1234.00$\pm$-1234.00 & 0.03$\pm$0.00 & reverb&1 & \cite{barth11} \\ 
\enddata
\tablecomments{(1) Name. (2) Distance. (3) BH mass. (4) Effective stellar velocity dispersion. (5) Method used for BH mass measurement. BH masses measured using stellar dynamics, gas dynamics, reverberation mapping and water masers are denoted by keywords star, gas, reverb, maser, respectively. (6) $1\sigma_s$, $2\sigma_s$ or $3\sigma_s$ confidence level on $M_\bullet$ uncertainty. (7) Literature reference. It is worth noting that $\cdots$ corresponds to missing data in the literature.}
\label{table:excluded}
\end{deluxetable*}
\end{longrotatetable}

\newpage


\bibliography{hurdle}{}
\bibliographystyle{aasjournal}



\end{document}